\documentclass[letterpaper]{article} 
\usepackage{aaai24}  
\usepackage{times}  
\usepackage{helvet}  
\usepackage{courier}  
\usepackage[hyphens]{url}  
\usepackage{graphicx} 
\usepackage{natbib}  
\usepackage{caption} 
\usepackage{algorithm}
\usepackage{algorithmic}

\usepackage{newfloat}
\usepackage{listings}
\DeclareCaptionStyle{ruled}{labelfont=normalfont,labelsep=colon,strut=off} 
\floatstyle{ruled}
\newfloat{listing}{tb}{lst}{}
\floatname{listing}{Listing}
\usepackage{amsmath}
\usepackage{amssymb}
\usepackage{mathtools}
\usepackage{amsthm}
\usepackage{dsfont}

\theoremstyle{plain}
\newtheorem{theorem}{Theorem}[section]
\newtheorem{proposition}[theorem]{Proposition}
\newtheorem{lemma}[theorem]{Lemma}
\newtheorem{corollary}[theorem]{Corollary}
\theoremstyle{definition}
\newtheorem{definition}[theorem]{Definition}

\theoremstyle{remark}

\DeclareMathOperator{\EX}{\mathbb{E}}

\newcommand\numberthis{\addtocounter{equation}{1}\tag{\theequation}}

\title{Peer Neighborhood Mechanisms:\\A Framework for Mechanism Generalization}
\author{
    Adam Richardson,
    Boi Faltings
}
\affiliations{
    École Polytechnique Fédérale de Lausanne\\
    Artificial Intelligence Laboratory
}

\begin{document}
\nocopyright
\maketitle

\begin{abstract}
Peer prediction incentive mechanisms for crowdsourcing are generally limited to eliciting samples from categorical distributions. Prior work on extending peer prediction to arbitrary distributions has largely relied on assumptions on the structures of the distributions or known properties of the data providers. We introduce a novel class of incentive mechanisms that extend peer prediction mechanisms to arbitrary distributions by replacing the notion of an exact match with a concept of neighborhood matching. We present conditions on the belief updates of the data providers that guarantee incentive-compatibility for rational data providers, and admit a broad class of possible reasonable updates.
\end{abstract}

\section{Introduction}
With the advent of machine learning in data analysis, governance, recommendation systems, and commercial products, there is an increasing need for abundant and accurate data with which to train these models. In some contexts, acquiring the necessary data can be expensive, slow, or require access to private information like medical records. For this reason, there has been ample recent research on \emph{crowdsourcing} for data acquisition. The primary goal of such research is to design \emph{incentive mechanisms}, or payment schemes, that incentivize independent, rational Agents to provide accurate data to a Center.

For many years, the gold standard for such incentive mechanisms has been the class of \emph{Peer prediction} mechanisms. Peer prediction mechanisms operate in the absence of any a priori baseline evaluation of the quality of reports. Generally, a Peer prediction mechanism works by comparing an Agent's report to a randomly selected report from another Agent, called a Peer, during the same data collection period. By examining correlations between reports, such mechanisms can induce desirable equilibria. In most settings, the Center wants the Agents to truthfully report some observation about a real world phenomenon. The major disadvantage of such mechanisms that they are generally only applicable for eliciting data from categorical distributions because they rely on a notion of report matching, i.e. the Agent and Peer report a sample from the same category. We refer to such mechanisms as \emph{Peer Consistency} mechanisms. Because they are restricted to categorical distributions, they eschew any notion of locality among the categories. This disadvantage presents a clear theoretical roadblock for applying such mechanisms to arbitrary distributions, since continuous random variables are only understandable through measuring local neighborhoods.

Our work presents a novel framework for extending Peer Consistency mechanisms to arbitrary distributions. We call such extensions \emph{Peer Neighborhood} mechanisms. To our knowledge, this is the only work that does so without assuming that the Center possesses a priori knowledge of properties of the Agents or of the underlying distribution. We only assume that Agents are rational and that they follow some reasonable belief update conditions, which we will show admit a broad class of updates. By analyzing an extension of the Peer Truth Serum, we prove that it can admit truthful Bayes-Nash Equilibria on the ex-ante game produced by the mechanism.

\begin{figure}[t!]
\centering
\centerline{\includegraphics[width=0.9\columnwidth]{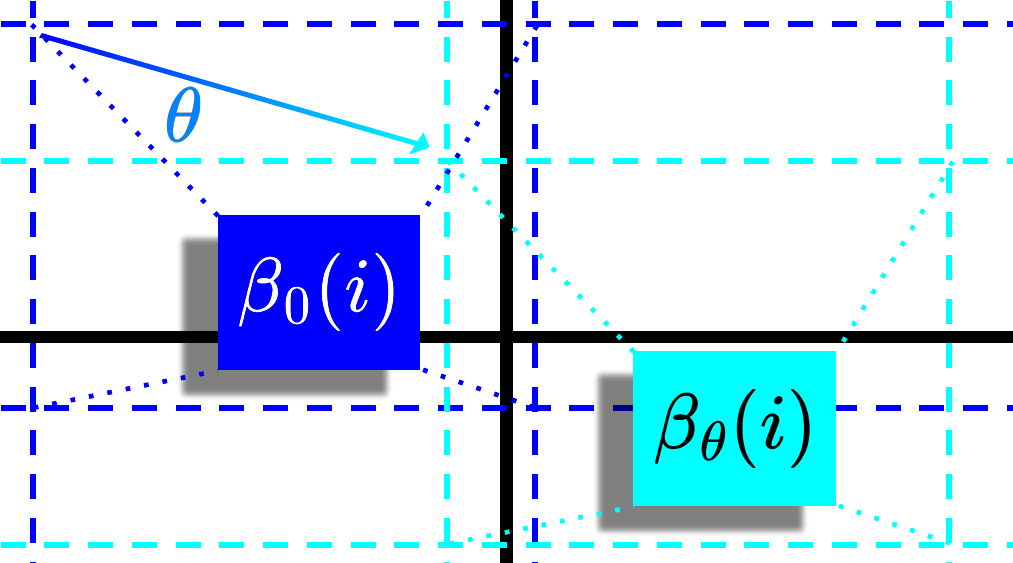}}
\caption{An example of a partition family on $\mathbb{R}^2$ with $\theta$ representing translations of the bins: $\beta_0(i)$ transforms into $\beta_\theta(i)$. Partition families are used to construct Peer Neighborhoods.}
\label{fig:bins_diagram}
\end{figure}

\subsection{Related Work}
Our work builds on a long line of advancements in incentive mechanism design based on Peer Consistency. The most basic such mechanism is Output Agreement, which simply pays Agents a constant reward when their reports are identical \cite{von2004labeling, waggoner2014output}. The class of Peer Consistency mechanisms was later broadened with the concept of Proper Scoring Rules \cite{winkler1996scoring, miller2005eliciting}. Other notable examples are multi-task peer prediction mechanisms, such as the Correlated Agreement mechanism \cite{dasgupta2013crowdsourced, shnayder2016informed}. There is also significant literature on Bayesian mechanisms, starting in economics \cite{d1979incentives, mcafee1992correlated} with trying to elicit private, correlated information. Subsequently, the Bayesian Truth Serum was developed for eliciting subjective private information \cite{prelec2004bayesian}. The distinguishing feature of these Bayesian mechanisms compared to classical Peer Consistency is the requirement that the reports include not merely a data sample, but also the Agent's estimate of the probability of the sample. While the additional information provided to the Center has allowed for extensions such as Robust BTS \cite{witkowski2012robust, radanovic2014incentives}, but it comes at great additional cost in practical terms and in the complexity of the mechanism. For this reason, we focus on Peer Consistency mechanisms which are \emph{minimal}, meaning that they do not require Agents to report anything other than the sample.

We primarily build on the work of Radanovic et al. on the Peer Truth Serum (PTS), which expands on the Output Agreement mechanism by considering different update conditions for Agent beliefs \cite{radanovic2016incentives}. The Output Agreement mechanism can be characterized as incentive-compatible with respect to the \emph{self-dominating} update condition. The PTS is incentive-compatible with respect to a much broader class of updates, characterized by the \emph{self-predicting} condition. Like the Peer Consistency mechanisms before it, the PTS is only applicable to categorical distributions. It has long been a goal of incentive mechanism design to uncover general concepts that allow for application over arbitrary distributions. One attempt is the Logarithmic Peer Truth Serum (LPTS), which does so by assuming a locality structure among the Agents so that more localized Peers have more similar statistical properties \cite{radanovic2015incentivizing}, and the Personalized Peer Truth Serum, which extends the LPTS for subjective private data \cite{goel2020personalized}. Finally, the work of \cite{chen2020truthful} and \cite{kong2019information} consider mechanisms which reward Agents according to the mutual information between reports, but assume that the distribution can be parameterized as an element of a known family of distributions. We see that in all cases, the ability to extend Peer Consistency concepts to arbitrary distributions relies on a priori assumptions about the structure of the distributions or the Agent properties. Our work eliminates these assumptions by considering a novel mechanism extention concept, which we call Peer Neighborhoods, and identifying a class of Agent updates that make such mechanisms incentive-compatible.

\section{Model}
In a crowdsourcing setting there is a Center that wishes to learn an arbitrary distribution $\Phi$, which we call the \emph{true distribution}, but the Center can't probe this distribution in a meaningful way. The Center tries to learn the distribution by collecting \emph{reports} from a set of independent, self-interested Agents who can sample $\Phi$ to produce an \emph{observation}. Because the Agents are self-interested, they must be \emph{incentivized} to produce reports that help the Center learn $\Phi$. The incentive an Agent experiences is a personal utility function that depends on the Agent's reporting strategy and a set of \emph{beliefs} the Agent has about the setting, such as the distribution $\Phi$ and the reporting strategies of other Agents. Agents always act rationally, so they will adopt the reporting strategy which maximizes their expected utility under their current beliefs. The Center's goal is to choose a payment function which dispenses utility to the Agents in exchange for reports, such that Agents will be incentivized to adopt "good" reporting strategies. In our case, we seek \emph{truthful} reporting, meaning Agents report their observations.

In the setting we consider, the Agent does not have a static set of beliefs. We refer to the belief of the Agent about the true distribution before making the observation as $\pi$, the \emph{prior belief}. After making an observation $o$, the prior is updated to $\pi_o$, the \emph{posterior belief}. Prior to the data collection period, the Center also has its own belief about the true distribution, $R$, which it makes public to the Agents. We refer to $R$ as the \emph{public prior}. It is assumed that these probability measures are on a shared measurable space $(\Omega, \Sigma)$. When discussing arbitrary distributions, we will assume that $\Omega = \mathbb{R}^d$ for some $d$, and $\Sigma = \mathcal{B}(\Omega),$ the Borel sets of $\Omega$.

The game sequence is as follows: a data collection period begins with a set of Agents possessing prior beliefs. Each Agent then makes an observation and updates their prior to a posterior belief. Each Agent then makes a report to maximize its expected payment according to the mechanism, which is public knowledge. At the end of the data collection period, the Center will have received a set of reports $\{r\}$. For each report, it randomly picks a Peer report and performs some comparison between the two reports. This informs the payment for the report.

\section{Peer Neighborhood Mechanisms}
\subsection{Peer Consistency}
In the original setting for Peer Consistency mechanisms, $\Phi$ is a categorical distribution, and the mechanisms pay an Agent when its report matches with a randomly selected Peer report. We formalize this concept:
\begin{definition}
A \emph{Peer Consistency} mechanism is a mechanism which assumes some public prior $R$ with categorical distribution, takes a report $r$ from an Agent and a report $rr$ from a randomly chosen Peer, and pays the Agent $\tau_R(r, rr) = f(rr) + s_R(r)*\mathds{1}_{r=rr}$ where $f$ depends only on $rr$, and $s_R$ is a non-negative \emph{scoring function} which depends on the distribution $R$.
\end{definition}
When discussing the incentives of a Peer Consistency mechanism, this is typically in regards to some \emph{update condition}:
\begin{definition}
Given a prior probability measure $\pi$, an observation $o$, and a posterior probability measure $\pi_o$, an \emph{update condition} is a predicate $S(\pi, \pi_o)$. We call $S^*(\pi, \pi_o)$ the \emph{natural update condition} for some scoring function $s_\pi$ as $\forall x \ne o:$ $\pi_o(o)*s_\pi(o) > \pi_o(x)*s_\pi(x)$.
\end{definition}
A notable example that we will use is the \emph{self-predicting} update condition for the Peer Truth Serum \cite{radanovic2016incentives}. This is the natural update condition and is given by $s_\pi(r) = \frac{1}{\pi(r)}$, so the condition is $\forall x \ne o:$ $\frac{\pi_o(o)}{\pi(o)} > \frac{\pi_o(x)}{\pi(x)}$. Assumptions about Agent behaviors can be embedded in update conditions in a way that reflects an informal notion of probabilistic reasonableness. For example, the self-predicting condition satisfies Bayes's Rule.
\begin{definition}
Given a prior probability measure $\pi$ and an observation $o \in \Omega$, an \emph{update process} is a function $\mathcal{U}(\pi, o) = \pi_o.$ We say an update process satisfies an update condition $S$ if $\forall \omega \in \Omega: S(\pi, \mathcal{U}(\pi, \omega))$ is true.
\end{definition}
\begin{definition}
A Peer Consistency mechanism with public prior $R$ is \emph{incentive-compatible} with respect to an update condition $S$ if an Agent, with prior $\pi = R$ and an update process which satisfies $S$, believes that for any observation $o$, their expected payment, with the Peer reports distributed according to the posterior, is maximized by truthfully reporting $o$.
\end{definition}
\begin{proposition}\label{prop:PC_IC_NUC}
A Peer Consistency mechanism is incentive-compatible with respect to the natural update condition.
\end{proposition}
The proof is in Appendix~\ref{App:prop:PC_IC_NUC}.

\subsection{Partition Spaces}
Peer Neighborhood mechanisms place a layer of abstraction on top of Peer Consistency mechanisms to introduce a notion of locality. They do so by considering a \emph{family of partitions} of the space of reports. A standard approach to applying a Peer Consistency mechanism to a continuous distribution, such as a Gaussian distribution, is to pick some fixed discretization, or partition. Each \emph{bin} of the partition corresponds to a category for the mechanism. A mechanism with a truthful equilibrium for some update condition over a discrete distribution would then have a bin-truthful equilibrium over this continuous distribution if the Agent's belief update satisfies the update condition with respect to the bin-categories. An Agent would have an equal incentive to report any value inside the bin containing the truthful report.

By considering a family of partitions rather than a single one, the incentives can be refined. Consider a Gaussian distribution partitioned into integer length bins $(n, n+1]$. An incentive-compatible Peer Consistency mechanism would then incentivize any report with the correct integer value. If a second partition is introduced with bins $(n+\frac{1}{2}, n+\frac{3}{2}]$, and the Agent satisfies the update condition for both partitions, the report is now incentivized to be within an interval of length $\frac{1}{2}$, corresponding to the intersection of the truthful bins from each partition. We will see that if the partition family is constructed correctly, this intersection can be refined to contain only the truthful report.
\begin{definition}
A \emph{partition family} $T$ is a function which maps a parameter $\theta \in \Theta$ to a partition, which is a countable set of measurable bins $\beta$ that are disjoint and cover $\Omega$:
$T(\theta) = \{ \beta_\theta(i) \}_{i \in \mathbb{Z}_\theta^*}$ where $\mathbb{Z}_\theta^* \subseteq \mathbb{Z}$ and $\beta_\theta(i) \in \mathcal{B}(\Omega)$ such that $\forall \theta$, $\bigcup_{i \in \mathbb{Z}_\theta^*} \beta_\theta(i) = \Omega$ and $\forall i \ne j$,  $\beta_\theta(i) \cap \beta_\theta(j) = \emptyset$
\end{definition}
A simple example of a partition family over $\mathbb{R}^2$ is shown in Figure~\ref{fig:bins_diagram}. The Center must have some way of selecting a partition from the family. We have the Center pick the partition randomly according to some distribution, which we call the \emph{partition selection distribution}.
\begin{definition}
The \emph{partition selection distribution} is given by a probability measure $\Psi$ over some measurable space $(\Theta, \Sigma)$, where $\Theta$ is the set of parameters for the partition family and $\Sigma$ is some $\sigma$-field over $\Theta$. Without loss of generality, let $\Psi$ be supported on $\Theta$. We call the pair $(T, \Psi)$ the \emph{partition space}.
\end{definition}

For ease of reading we will often use the \emph{bin selection function} to identify the bin that contains a particular point:
\begin{definition}
The \emph{bin selection function} with respect to a partition family $T(\theta)$ is a function $\mathds{X}_\theta: \Omega \rightarrow \mathbb{Z}_\theta^*$ such that $\mathds{X}_\theta(z) = i$ if and only if $z \in \beta_\theta(i)$.
\end{definition}
The Bin Selection Function is well-defined as a result of the bins of each partition being disjoint and covering $\Omega$. It is important for the Center's implementation that this function be computable.

If an Agent is to have a strictly truthful incentive, there must be a non-zero probability of any other report failing to match with the truthful report under the partition family. We call this \emph{point-isolating}:
\begin{definition}
A partition space $(T, \Psi)$ is \emph{point-isolating} over $R$ if: $\omega_1 \ne \omega_2$ in the support of $R$ $\Rightarrow \Psi(\{\theta:\mathds{X}_\theta(\omega_1) \ne \mathds{X}_\theta(\omega_2)\}) > 0$.
\end{definition}
Reports with a matching probability of 0 in $R$ can result in infinite payments under some Peer Consistency mechanisms, such as the Peer Truth Serum. To avoid degenerate payments, we impose the following condition on the partition family:
\begin{definition}
A partition space $(T, \Psi)$ is \emph{bin-supported} over $R$ if $\forall \omega \in \Omega: \Psi(\{\theta: R(\beta_\theta(\mathds{X}_\theta(\omega)))=0\}) = 0$.
\end{definition}
In simpler terms, for any possible report, the probability of selecting a partition with a 0 $R$-probability bin containing the report is 0 in $\Psi$.
\begin{proposition}\label{prop:Exists_PI_BS}
$\forall R$, $\exists (T, \Psi)$ such that $(T, \Psi)$ is point-isolating and bin-supported over $R$
\end{proposition}
The proof is in Appendix~\ref{App:prop:Exists_PI_BS}.

\subsection{The Mechanism Extension}
Now, we can introduce the Peer Neighborhood mechanism extension. First we must modify the probability measures:
\begin{definition}
Let $\pi$ be a probability measure on $(\Omega, \mathcal{B}(\Omega))$. Let $T(\theta)$ be a partition of $\Omega$. Then for $i \in \mathbb{Z}^*_\theta$, let the \emph{partitioned probability measure} $\pi^\theta(i) = \pi(\beta_\theta(i))$.
\end{definition}
We can then extend any Peer Consistency mechanism as follows:
\begin{definition}
Given some Peer Consistency mechanism with payment function $\tau$, we define the bin-extension payment function with respect to some partition $T(\theta)$ as $\tau^\theta_R(r, rr) = \tau_{R^\theta}(\mathds{X}_\theta(r), \mathds{X}_\theta(rr)).$ Then given some partition selection distribution $\Psi$ such that $(T, \Psi)$ is point-isolating and bin-supported over $R$, the \emph{Peer Neighborhood extension mechanism} pays according to:
\begin{small}
\begin{equation}
\tau^\Psi_R(r, rr) = \EX_{\theta \sim \Psi}[\tau^\theta_R(r, rr)]
\end{equation}
\end{small}
\end{definition}

\subsection{Incentive-Compatibility}
Given some Peer Consistency mechanism with scoring function $s$, we wish to discover an update condition $S^{(T,\Psi)}$ for which the associated Peer Neighborhood extension mechanism is incentive-compatible. We suggested earlier that the incentivized report region could be refined as long as the Agent is incentivized to be bin-truthful for all the partitions, so the most straightforward condition is that $S$ is satisfied with probability 1 in $\Psi$.
\begin{definition}\label{def:PI_condition}
Given a prior $\pi$, a posterior $\pi_o$, a partition space $(T, \Psi)$, and a Peer Consistency mechanism with scoring function $s$, the \emph{Partition-Invariant} (PI) update condition $S^{(T,\Psi)}_{PI}$ takes the form:
\begin{small}
\begin{equation}
\Psi(\{\theta: S^*(\pi^\theta, \pi^\theta_o)\})=1
\end{equation}
\end{small}
\end{definition}
\begin{proposition}\label{prop:IncentiveComp_PI}
Given a Peer Consistency mechanism with payment function $\tau$ and scoring function $s$, and given $(T, \Psi)$ point-isolating over $R$, the Peer Neighborhood extension mechanism $\tau^\Psi_R(r, rr)$ is incentive-compatible with respect to the update condition $S^{(T,\Psi)}_{PI}$.
\end{proposition}
The proof is in Appendix~\ref{App:prop:IncentiveComp_PI}. While this update condition clearly guarantees incentive-compatibility of the Peer Neighborhood extension mechanism, we will see that it is stronger than necessary. We present a more relaxed update condition:
\begin{definition}\label{def:PE_condition}
Given a prior $\pi$, a posterior $\pi_o$, a partition space $(T, \Psi)$, and a Peer Consistency mechanism with scoring function $s$, the \emph{Partition-Expected} (PE) update condition $S^{(T,\Psi)}_{PE}$ takes the form:
\begin{small}
\begin{align*}
\forall x \ne o: &\EX_{\theta \sim \Psi}[\pi^\theta_o(\mathds{X}_\theta(o))*s_{\pi^\theta}(\mathds{X}_\theta(o))] \\
> &\EX_{\theta \sim \Psi}[\pi^\theta_o(\mathds{X}_\theta(x))*s_{\pi^\theta}(\mathds{X}_\theta(x))]
\end{align*}
\end{small}
\end{definition}
\begin{proposition}\label{prop:IncentiveComp_PE}
Given a Peer Consistency mechanism with payment function $\tau$ and scoring function $s$, and given $(T, \Psi)$ point-isolating over $R$, the Peer Neighborhood extension mechanism $\tau^\Psi_R(r, rr)$ is incentive-compatible with respect to the update condition $S^{(T,\Psi)}_{PE}$.
\end{proposition}
The proof is in Appendix~\ref{App:prop:IncentiveComp_PE}. Furthermore, we show that the PE condition is a relaxed form of the PI condition, in that any update process which satisfies PI also satisfies PE:
\begin{lemma}\label{lem:PI_implies_PE}
Given a partition space $(T, \Psi)$ that is point-isolating over $R$, and a Peer Consistency mechanism with scoring function $s$, any update process $\mathcal{U}(R, o)$ which satisfies the PI extended update condition $S^{(T,\Psi)}_{PI}$ also satisfies the PE extended update condition $S^{(T,\Psi)}_{PE}$.
\end{lemma}
The proof is in Appendix~\ref{App:lem:PI_implies_PE}.

\section{Analysis of Update Processes}
We have constructed a framework that extends Peer Consistency mechanisms to arbitrary distributions, but the crux of this extension is the Partition-Expected update condition, which is necessarily more restrictive than the natural update condition for the underlying discrete mechanism. We will examine what types of update processes satisfy this condition, but we must first address a practical concern which will further restrict update processes, namely whether or not an update process is consistent with convergence of the posterior to the true distribution.

\subsection{Update Convergence}
When an Agent makes on observation and computes a posterior according to some update process, that process should generally bring the Agent's belief closer to the true distribution. With finite observations, it is always possible that an Agent can observe a very unlikely sequence, leading to a bias in the posterior. But in the limit of infinite observations, the posterior should converge to the true distribution. We then wish to describe update processes which can be performed iteratively to converge to the true distribution.
\begin{definition}
Consider a sequence of update processes $\mathcal{U}_i$ for all $i \in \mathbb{Z}_+$. The sequence is \emph{convergent} if, when the sequence $\{\mathcal{U}_i\}$ is applied iteratively to a sequence of i.i.d. observations $\{o_i\}$ sampled from the true distribution, the sequence of posteriors converges in distribution to the true distribution.
\end{definition}
In order to get a better grasp on such update processes, we will restrict ourselves to a particular type of update process, which we call \emph{additive}:
\begin{definition}
An update process $\pi_o = \mathcal{U}(\pi, o)$ is \emph{additive} if $\pi_o = (1-\alpha)\pi + \alpha K_o$ where $K_o$ is a probability measure which we call the \emph{update kernel}, and $\alpha \in (0,1)$.
\end{definition}
Often we will refer to an update process of this form simply by referring to the update kernel. The Agent picks $(1-\alpha)$ to represent the Agent's confidence in the accuracy of its prior.

Suppose an Agent were to observe a sequence of i.i.d. samples from the true distribution and sequentially update. It would be unreasonable for the Agent to update to a different posterior depending on the order of the sequence of samples, so the update kernels should be given equal weight. Given some additive update with $\alpha_1$, the next update must then have $\alpha_2 = \frac{\alpha_1}{1+\alpha_1}$. A simple choice for $\alpha_1$ would be $\frac{1}{k}$ for some positive integer $k$, so $\alpha_2 = \frac{1}{k+1}$. This process of decreasing $\alpha$ like $\frac{1}{n}$ can be applied iteratively, and we refer to this as an \emph{additive update sequence}:
\begin{definition}
An \emph{additive update sequence} is given by $\mathcal{U}_k(\pi, o) = \frac{k}{k+1}\pi + \frac{1}{k+1}K_o$. For a sequence of $n$ observations $\{o_i\}$, the final posterior is given by: $\pi_{\{o_i\}} = \mathcal{U}_{k+n-1}(\mathcal{U}_{k+n-2}(\dots \mathcal{U}_k(\pi, o_1)\dots , o_{n-1}), o_n) = \frac{k}{k+n}\pi + \frac{1}{k+n}\sum_{i=1}^n K_{o_i}$.
\end{definition}
First we show that the convergence of this update sequence depends only on the update kernels. We say this update sequence is \emph{prior agnostic}:
\begin{lemma}\label{lem:prior_agnostic}
The additive update sequence converges in distribution to the same distribution as the average of the kernels: $\frac{1}{n}\sum_{i=1}^n K_{o_i} \xrightarrow[]{d} X \iff \frac{k}{k+n}\pi + \frac{1}{k+n}\sum_{i=1}^n K_{o_i} \xrightarrow[]{d} X$.
\end{lemma}
The proof is in Appendix~\ref{App:lem:prior_agnostic}. We now address the structure of the update kernel $K$. We will consider two types of kernels, the first is a simple point mass. We call this the \emph{Empirical Update}:
\begin{definition}
The \emph{Empirical Update} is the additive update where $K_o(A) = \mathds{1}_{o \in A}$.
\end{definition}
\begin{proposition}\label{prop:EmpUpdate_converge}
The Empirical Update sequence is convergent.
\end{proposition}
The proof is in Appendix~\ref{App:prop:EmpUpdate_converge}. The second type of kernel we wish to address is a kernel with a continuous CDF. We call such updates \emph{continuous}:
\begin{definition}
An additive update is \emph{continuous} if the CDF $F_K$ of the update kernel $K$ is continuous.
\end{definition}
We show that an additive continuous update sequence is convergent if the sequence of kernels satisfies a condition on the partial sums of their concentrations around the observed samples:
\begin{theorem}\label{thm:ConcentrationConverge}
Let $O_n = \{o_i\}_{i \in [1,n]}$ be a sequence of i.i.d random variables distributed with CDF $F(x)$. Let $K_{o_i}$ be a continuous update kernel. Define $H_n(x) = \frac{1}{n} \sum_{i=1}^n F_{K_{o_i}}(x)$. Consider the random variables $Y_i = |X_i - o_i|$ where $X_i$ is distributed according to $K_{o_i}$. Define $C_i(\epsilon) = P(Y_i \ge \epsilon)$. If $\forall \epsilon > 0: \lim_{n \rightarrow \infty} \frac{1}{n}\sum_{i=1}^{n} C_i(\epsilon) =0$, then $H_n \xrightarrow[]{d} F$.
\end{theorem}
The proof is in Appendix~\ref{App:thm:ConcentrationConverge}. The convergence of the additive continuous update sequence with kernels satisfying this condition follows directly from Lemma~\ref{lem:prior_agnostic}.

We present a very simply condition on the kernels which will satisfy Theorem~\ref{thm:ConcentrationConverge}. The kernels merely need to have bounded support, with that bound converging to $0$.
\begin{corollary}\label{cor:BoundedConvergence}
Let $\Delta_i = \langle \delta_{i,1}, \delta_{i,2}, \dots, \delta_{i,d} \rangle$ with $\delta_{i,j} > 0$ and $\lim_{i \rightarrow \infty} \delta_{i,j} = 0$. Let $A_i = [o_i - \Delta_i, o_i + \Delta_i]$ Suppose $K_{o_i}(A) = 1$. Then an additive continuous update sequence with these kernels is convergent.
\end{corollary}
The proof is in Appendix~\ref{App:cor:BoundedConvergence}.

\subsection{Satisfying the Update Conditions}
We now analyze update processes which satisfy our extended update conditions PI and PE. Whether or not these conditions are satisfied depends heavily on the choice of scoring function $s$. We choose to focus our attention to the Peer Neighborhood extension of the PTS, commonly considered the canonical example of a Peer Consistency mechanism. We will then refer to the extension as the \emph{Peer Truth Neighborhood Extension} mechanism.
\begin{definition}
The \emph{Peer Truth Neighborhood Extension} (PTNE) mechanism is the Peer Neighborhood extension of the Peer Consistency mechanism with scoring function $s_R(r) = \frac{c}{R(r)}$ where $c$ is a positive constant, known as the Peer Truth Serum.
\end{definition}
The natural update condition for the PTS is $\forall x \ne o: \frac{\pi_o(o)}{\pi(o)}>\frac{\pi_o(x)}{\pi(x)}$, known as the \emph{self-predicting} update condition.

We first prove that the Empirical Update satisfies PI for the PTNE, and therefore also PE:
\begin{theorem}\label{thm:EmpUpdate_PI}
Given a partition space $(T, \Psi)$ that is point-isolating and bin-supported over $R$, the Empirical Update process satisfies $S^{(T, \Psi)}_{PI}$ for the PTNE mechanism.
\end{theorem}
The proof is in Appendix~\ref{App:thm:EmpUpdate_PI}. The Empirical Update is perfectly reasonable and is used quite frequently in modeling, but an Agent may want to make a continuous update as they may be unsure that their measurement of the true distribution is precise. But with additive continuous updates, whether or not they satisfy our update conditions depends on the structure of the partition space. To simplify analysis, we focus on partition spaces with a high degree of symmetry, which we call \emph{regular}:
\begin{definition}
A \emph{regular} partition space $(T, \Psi)$ over a fundamental set $\Omega = \mathbb{R}^d$ is one in which each bin is a rectangular prism with side lengths $\{l_i\}_{i \in [1,d]}$, i.e. $\forall \theta \in \Theta,$ $T(\theta) = \{ \bigotimes_{i=1}^d [l_i*(n_i-\frac{1}{2})+\theta_i, l_i*(n_i+\frac{1}{2})+\theta_i) \forall n_i \in \mathbb{Z}\}$ and $\Theta = \bigotimes_{i=1}^d [0, l_i)$, with $\Psi$ uniform over $\Theta$.
\end{definition}
This partition space is clearly point-isolating over any $R$ as it is point isolating for all $\omega \in \Omega$. We will assume that $R$ is such that this partition space is bin-supported over $R$.

\subsection{Bin Edge Conditions}

Let us consider some regular partition space. Let each bin have dimensions $L = \langle l_1, l_2, \dots , l_d \rangle.$ To simplify the notation, we will say the set $[-L, L) = \bigotimes_{i=1}^d [-l_i, l_i)$ We define the Bin Function $B: \mathbb{R}^d \rightarrow \{0, 1\}$ to be:
\begin{small}
\begin{equation*}
B(\omega) = \begin{cases} 
1 & \omega \in [-\frac{L}{2}, \frac{L}{2}) \\
0 & \text{otherwise}
\end{cases}
\end{equation*}
\end{small}
so the Bin Function is just an indicator for a bin centered at $0$. Assume that an agent with prior and posterior $\pi$ and $\pi_o$ respectively has PDFs $f_{\pi}$ and $f_{\pi_o}$. It's not necessary that such PDFs exist, but we make this assumption for ease of presentation. Let us define the overhead $\sim$ to be the operator such that for a function $f$, $\widetilde{f}(x) = (f \circledast B)(x)$, where $\circledast$ is the convolution operator. Then $\widetilde{f}_{\pi}(x)$ is just the prior probability of a sample landing in a bin centered at $x$, and same for the posterior $\widetilde{f}_{\pi_o}(x)$. These functions can be computed only using CDFs, but it is simpler to define them this way. The quantities we are concerned with regarding the PI and PE conditions for the PTNE mechanism are the ratios $Q(x) = \dfrac{\widetilde{f}_{\pi_o}(x)}{\widetilde{f}_{\pi}(x)}$. The expected payment for reporting $x$ is simply $\widetilde{Q}(x)$. If the update process is additive continuous, then $Q$ and $\widetilde{Q}$ are continuous.

\begin{figure}[t!]
    \centering
    \includegraphics[width=0.9\linewidth]{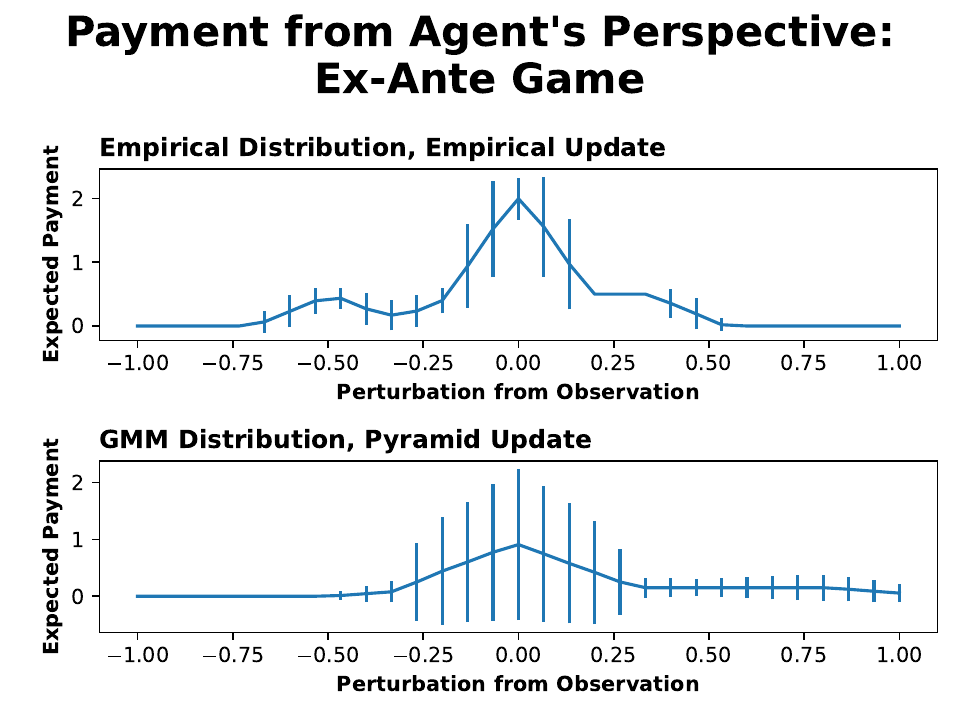}
    \includegraphics[width=0.9\linewidth]{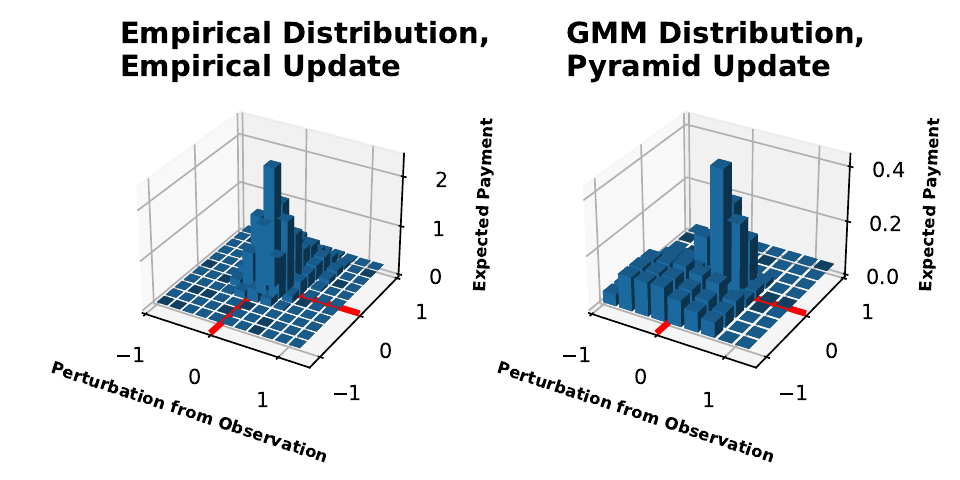}
    \caption{Expected payments for reports perturbed from the observation, computed over an Agent's posterior. Error bars are one standard deviation. In the 2D figures, red lines show the location of the maximum expected payment.}
    \label{fig:payments_agent}
\end{figure}

The PI condition gives us the following constraints: Let $N = \langle n_1, n_2, \dots, n_d \rangle$ where $n_i \in \mathbb{Z}$ and $N \ne 0$. Then $\forall x \in (o - \frac{L}{2}, o + \frac{L}{2}]: Q(x) > Q(x + N \ast L)$ where $\ast$ is element-wise multiplication. Let $Q_o(x) = Q(o+x)$. From the continuity of $Q$, it follows that for all $i \in [1, d]$ and all $\delta_i \in [-\frac{l_i}{2}, \frac{l_i}{2}]$:
\begin{small}
\begin{equation}\label{PIConditions}
Q_o(\delta_1, \dots , -\frac{l_i}{2}, \dots , \delta_d) = Q_o(\delta_1, \dots , \frac{l_i}{2}, \dots , \delta_d)
\end{equation}
\end{small}
We see that these are equalities on every pair of opposing points on the boundary of the bin centered at $o$.

The PE condition simply constrains $o$ to be the global maximum of $\widetilde{Q}$. As long as the continuous update kernel has mass at $o$ and has sufficiently bounded support, if $o$ is a local maximum of $\widetilde{Q}$, then it will be the global maximum. We'll discuss what sufficiently bounded means later. We write the PE constraint:
\begin{small}
\begin{equation*}
\nabla_x \widetilde{Q} |_{x=o} = 0, \qquad \nabla^2_x \widetilde{Q} |_{x=o} < 0.
\end{equation*}
\end{small}
From the continuity of $\widetilde{Q}$ we obtain conditions that are much less restrictive than for PI. Let $L_{-i}$ be the vector $L$ with entry $l_i$ removed, and $\Delta_i$ be the vector of $\delta_j$s with entry $\delta_i$ removed. Then $\forall i \in [1, d]$:
\begin{small}
\begin{align*}
&\int_{-\frac{L_{-i}}{2}}^{\frac{L_i}{2}} Q_o(\delta_1, \dots , -\frac{l_i}{2}, \dots , \delta_d)\, d \Delta_i \\
= &\int_{-\frac{L_{-i}}{2}}^{\frac{L_i}{2}} Q_o(\delta_1, \dots , \frac{l_i}{2}, \dots , \delta_d)\, d \Delta_i \numberthis \label{PEConditions}
\end{align*}
\end{small}
We see that rather than having an equality for every pair of opposing points on the boundary of the bin centered at $o$, we have a single equality for each opposing boundary surface of the bin. This is equivalent to the constraints for PI in one dimension, since the opposing boundary surfaces are just a single pair of points, but in higher dimensions it is much less constraining.

We also see that a continuous update kernel has "sufficiently bounded support" if it has support within $(o-L, o+L]$. From now on we will refer to such an update kernel as \emph{bin-bounded}.

\begin{figure}[t!]
    \centering
    \includegraphics[width=0.9\linewidth]{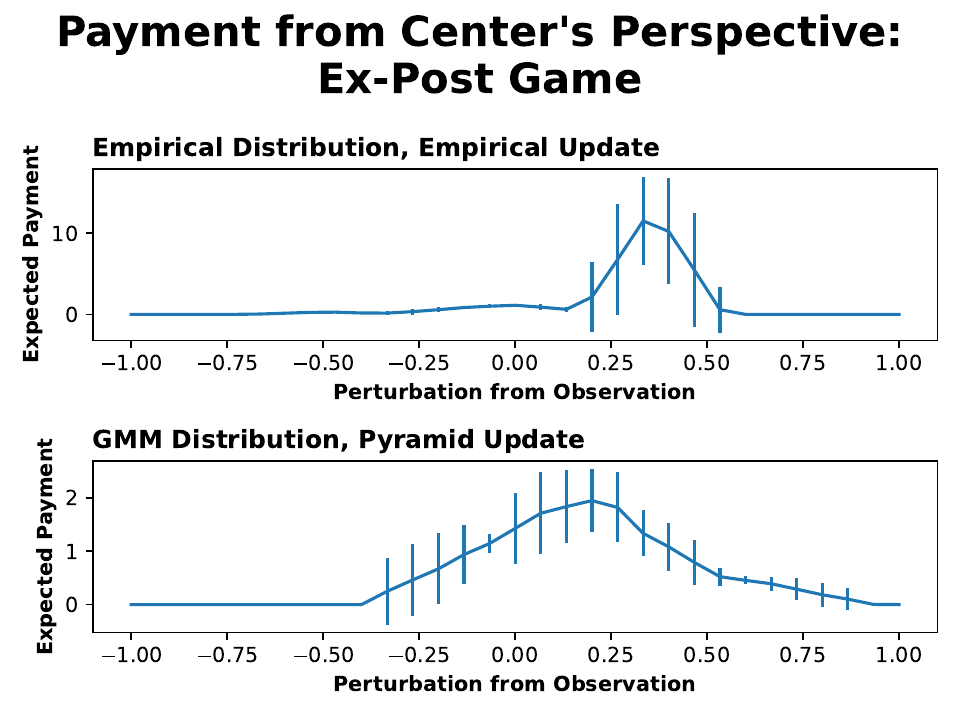}
    \includegraphics[width=0.9\linewidth]{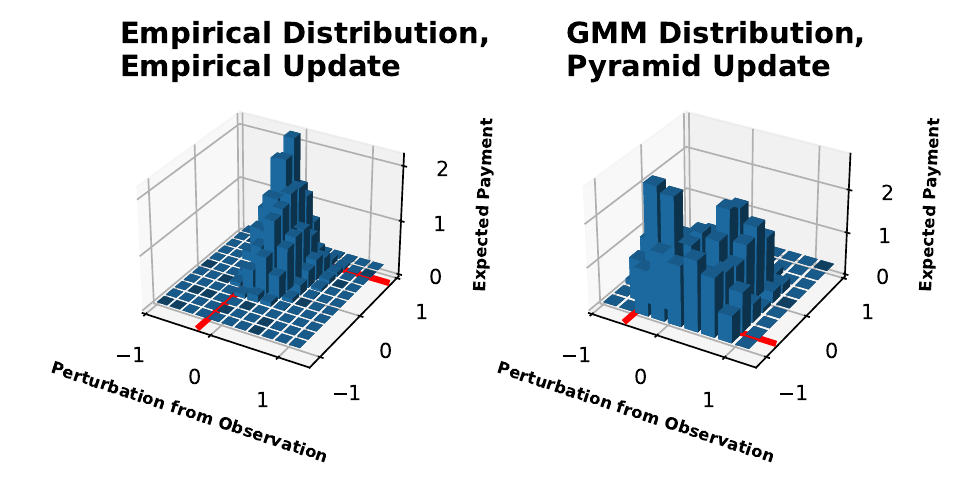}
    \caption{Expected payments for reports perturbed from the observation, computed over truthful Peer reports. Error bars are one standard deviation. In the 2D figures, red lines show the location of the maximum expected payment.}
    \label{fig:payments_center}
\end{figure}

\subsubsection{Failing PI}
We will first show that it is impossible in general for a bin-bounded continuous update kernel to satisfy both PI and for the associated additive update sequence to be convergent in dimensions higher than one. We will show the proof for two dimensions, but the same argument applies to higher dimensions.

\begin{lemma}\label{lem:BoundedCenterMass}
Given a regular partition space on $\mathbb{R}^2$, let each bin have dimensions $L = \langle l_1, l_2 \rangle$. Let $\Delta = \langle \frac{l_1}{2}, \frac{l_2}{2} \rangle$, and $A = [z-\Delta, z+\Delta]$. There is a prior $\pi$ such that a continuous update kernel must have bounded probability on $A$: $K_o(A) < x < 1$ in order to satisfy PI for the PTNE mechanism.
\end{lemma}
The proof is in Appendix~\ref{App:lem:BoundedCenterMass}. With this we can prove that a continuous update kernel cannot allow for an additive update sequence that is convergent:
\begin{theorem}\label{thm:PI_impossible}
Given a regular partition space on $\mathbb{R}^2$, there is a prior $\pi$ and true distribution $\Phi$ such that a continuous update kernel cannot satisfy PI for the PTNE mechanism and admit an additive update sequence that is convergent.
\end{theorem}
The proof is in Appendix~\ref{App:thm:PI_impossible}.

\subsubsection{Satisfying PE}
We will now show that it is always possible to construct a sequence of continuous update kernels that satisfy both PE and are convergent. We will construct these explicitly. First we will restrict our construction so that all the probability of the kernel is within a bounded region $A = [x - \Delta, x + \Delta]$ which contains the observation point $o$, and where $\Delta$ can be arbitrarily small. From Corollary~\ref{cor:BoundedConvergence}, we find that by allowing the sequence $\Delta_i$ to converge to 0, this update sequence will be convergent. We are left only to show that the kernel construction satisfies PE:
\begin{theorem}\label{thm:PEPyramids}
Given a regular partition space on $\mathbb{R}^d$, for any prior $\pi$, there exists a continuous update kernel that satisfies PE for the PTNE mechanism and is arbitrarily bounded around a point $o$.
\end{theorem}
The proof is in Appendix~\ref{App:thm:PEPyramids}. It involves constructing update kernels with PDFs as hyper-pyramids with a peak at $o$ and a base at $[x-\Delta, x+\Delta]$ for some arbitrary positive $\Delta<L$ where $L$ is the dimensions of the bins. The proof demonstrates the existence of $x$ such that $o \in [x-\Delta, x+\Delta]$ and the kernel satisfies the bin edge conditions in Equation~\ref{PEConditions}. The proof goes further by suggesting a method for computing this $x$, which we take advantage of in our simulations.

\section{Simulations}

\begin{figure}[t!]
\centering
\centerline{\includegraphics[width=0.9\columnwidth]{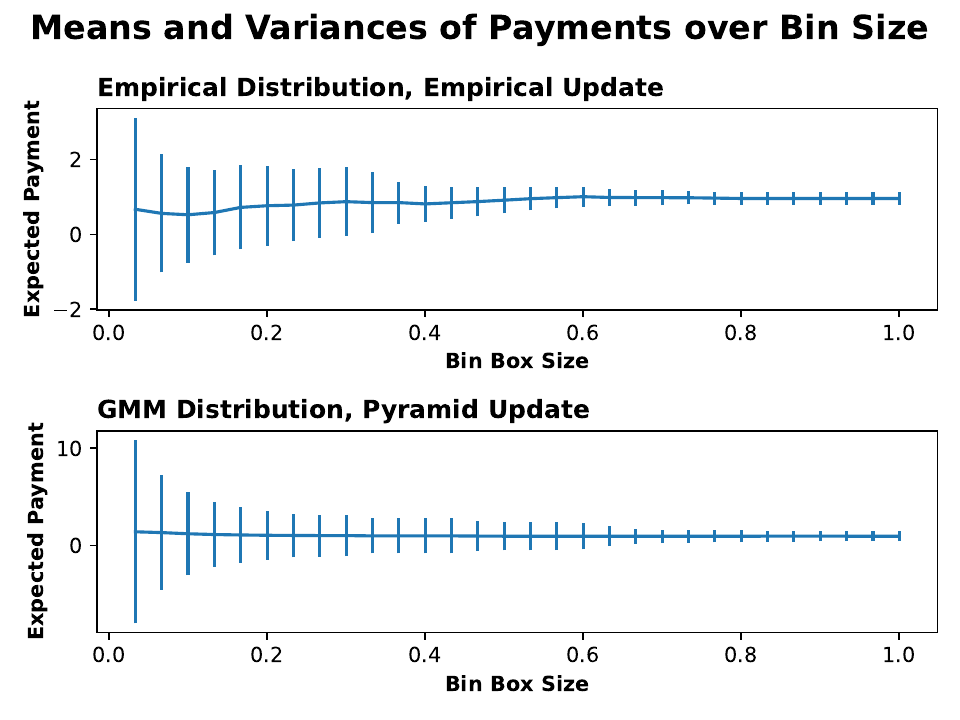}}
\caption{Smaller bins produce a larger variance in payments. Error bars are one standard deviation squared.}
\label{fig:boxes}
\end{figure}

We conduct simulations using the PTNE mechanism to demonstrate the accuracy and stability of the incentives in settings with finite data for constructing models and finite peer reports. We use artificially generated data to form the true and public distributions, which can then be used to analyze expected payments and actual payments from samples. We present two data models: 1) an Empirical distribution constructed by taking finite samples with randomized frequencies, and 2) a continuous distribution constructed as a weighted sum of Gaussian distributions, or a Gaussian Mixture Model (GMM). For the first model, Agents use the Empirical Update, while for the second they update using Pyramid kernels as described in the proof of Theorem~\ref{thm:PEPyramids}. In all cases, the partition space is regular. We provide details of the simulation parameters in Appendix~\ref{App:Simulations}.

\subsection{Report Perturbation}
We simulate the expected payments for an Agent reporting a point that is a perturbation of the observation, meaning the payment for the observation itself is at $0$. Figure \ref{fig:payments_agent} shows the expected payments computed from the perspective of the Agent over the posterior. The error bars show the standard deviation with respect to the Partition Selection distribution. We observe that the Agent believes their payment will be maximized by truthfully reporting the observation, as expected from the theory. Figure \ref{fig:payments_center} shows the same expected payments, but this time computed over a set of truthful Peer reports collected by the Center. The expected payment from the Center's side is not necessarily maximized at the observation point. Since the public distribution is different from the true distribution, the observation made by the Agent might be an over-represented point in the public distribution. If this is the case, the Agent will be underpaid when compared against Peers reporting samples from the true distribution, and some perturbation of the observation might pay better. One can visually inspect the true, public, and kernel distribution figures in Appendix~\ref{App:Simulations} to see how the relationships between them produce the skewed figures. This does not matter for the incentives in the ex-ante game that the Agents play, however, as it is an ex-post calculation.

\subsection{Payment Stability}
We simulate the expectation and variance of payments with respect to bin size for the partition. The bin size can affect the expected payment of the Agent in complicated ways when you take into account that bin-bounded kernels must account for the bin size. From the perspective of the Center, however, the bin size should not affect the expected payment. A smaller bin means a lower probability of matching, but a proportionately higher payment when matching. Intuitively, a smaller bin size will lead to higher variance in the payments. We demonstrate this relationship in Figure~\ref{fig:boxes}. The stability of the payments could be a consideration for designing the mechanism to take into account either Centers or Agents who aren't risk-neutral.

\section{Conclusion}
We present Peer Neighborhood mechanisms, a novel framework for extending Peer Consistency from discrete distributions to arbitrary distributions with minimal additional constraints or assumptions. We formulate the explicit extension of the Peer Truth Serum to the PTNE and prove its incentive-compatibility. We show that the strengthened Agent update condition, the Partition Expected extension, still admits a broad class of reasonable update processes. Finally, we conduct simulations to demonstrate the strength of the incentives with respect to perturbations from truthfulness, and the stability of payments with respect to the bin size of the partitions chosen by the Center. Peer Neighborhood mechanisms are not only practically implementable, but present a rich concept for future research on mechanism generalization.

\appendix

\section{Peer Neighborhood Mechanisms}

\subsection{Peer Consistency}

Proof of Proposition~\ref{prop:PC_IC_NUC}
\begin{proof}\label{App:prop:PC_IC_NUC}
Consider an Agent with prior $\pi = R$, observation $o$, and posterior $\pi_o$. Suppose the update satisfies $\forall x \ne o:$ $\pi_o(o)*s_\pi(o) > \pi_o(x)*s_\pi(x)$. Then $\forall x \ne o:$ $\EX_{rr \sim \pi_o}[s_R(o)*\mathds{1}_{o=rr}] > \EX_{rr \sim \pi_o}[s_R(x)*\mathds{1}_{x=rr}]\Rightarrow \forall x \ne o:$ $\EX_{rr \sim \pi_o}[\tau_R(o, rr)] > \EX_{rr \sim \pi_o}[\tau_R(x, rr)]$.
\end{proof}

\subsection{Partition Spaces}

Proof of Proposition~\ref{prop:Exists_PI_BS}
\begin{proof}\label{App:prop:Exists_PI_BS}
Suppose $\Omega = \mathbb{R}^d$, define $T^*(\theta) = \{ \forall n_j \in \mathbb{Z}: \bigotimes_{j=1}^d [n_j+\theta_j, n_j+1+\theta_j)\}$ and $\Theta = \bigotimes_{j=1}^d [0, 1)$. Let $\Psi$ be uniform over $\Theta$. For all $\theta$, we contract the partition $T^*(\theta)$ by merging any bin with probability 0 in $R$ with the closest bin with positive probability in $R$, breaking ties with any deterministic process. We call this new partition family $T$ and claim that $(T, \Psi)$ is point-isolating and bin-supported over $R$.

By construction, $(T, \Psi)$ is clearly bin-supported over $R$, as all bins with probability 0 have been merged with bins of positive probability. Suppose two points $\omega_1 \ne \omega_2$ are in the support of $R$. Then there is a set of $\omega$ that separates them in $T^*$ with probability $\epsilon>0$ in $\Psi$, but in order for $T$ to not be point isolating, the probability that the bin containing one of the points has 0 probability in $R$ must be $\epsilon$. The set of $\omega$ which places one of the points on the closed boundary of a bin is clearly probability 0 in $\Psi$, and for all other $\theta$ separating the two points there exist open sets around the two points that are contained within the bins, and there is a positive probability in $\Psi$ of picking partitions that separate the open sets. Because the points are in the support of $R$, those open sets have positive probability in $R$, so the bins both have positive probability, so they can't be merged.
\end{proof}

\subsection{Incentive Compatibility}

Proof of Proposition~\ref{prop:IncentiveComp_PI}
\begin{proof}\label{App:prop:IncentiveComp_PI}
Without loss of generality, let $f(rr) = 0$ in $\tau.$ Suppose $\tau^\Psi_R(r, rr)$ is not incentive-compatible with respect to $S^{(T,\Psi)}_{PI}$. Then there exists an update process $\pi_o = \mathcal{U}(R, o)$ which satisfies $S^{(T,\Psi)}_{PI}$, but the expected payment $\EX_{rr \sim \pi_o}[\tau^\Psi_R(o, rr)] < \EX_{rr \sim \pi_o}[\tau^\Psi_R(x, rr)]$ for some $x \ne o$. This implies that 
\begin{align*}
&\EX_{\theta \sim \Psi}[\EX_{rr \sim \pi_o}[\tau^\theta_R(o, rr)]] \\
< &\EX_{\theta \sim \Psi}[\EX_{rr \sim \pi_o}[\tau^\theta_R(x, rr)]] \\
\Rightarrow &\EX_{\theta \sim \Psi}[\pi^\theta_o(\mathds{X}_\theta(o))*s_{R^\theta}(\mathds{X}_\theta(o)] \\
< &\EX_{\theta \sim \Psi}[\pi^\theta_o(\mathds{X}_\theta(x))*s_{R^\theta}(\mathds{X}_\theta(x))]
\end{align*}
The partition space $(T, \Psi)$ being point-isolating implies that $\Psi(\{\theta: \mathds{X}_\theta(o) \ne \mathds{X}_\theta(x)\})>0$, so there is a set of $\theta$ with positive probability in $\Psi$ such that $\Psi(\{\theta: \pi^\theta_o(\mathds{X}_\theta(o))*s_{R^\theta}(\mathds{X}_\theta(o)) < \pi^\theta_o(\mathds{X}_\theta(x))*s_{R^\theta}(\mathds{X}_\theta(x))\})>0.$ It follows directly that for this set of $\theta$, $S^*(R^\theta, \pi^\theta_o)$ is false, violating our assumption that $\mathcal{U}(R, o)$ satisfies $S^{(T,\Psi)}_{PI}$.
\end{proof}

Proof of Proposition~\ref{prop:IncentiveComp_PE}
\begin{proof}\label{App:prop:IncentiveComp_PE}
This proof follows directly from the previous proof, in which we showed that if $\tau^\Psi_R(r, rr)$ is not incentive compatible with respect to $S^{(T,\Psi)}_{PE}$, then
\begin{align*}
\exists x \ne o: &\EX_{\theta \sim \Psi}[\pi^\theta_o(\mathds{X}_\theta(o))*s_{R^\theta}(\mathds{X}_\theta(o)] \\
< &\EX_{\theta \sim \Psi}[\pi^\theta_o(\mathds{X}_\theta(x))*s_{R^\theta}(\mathds{X}_\theta(x))]
\end{align*}
which violates the assumption that the update $\pi_o = \mathcal{U}(R, o)$ satisfies $S^{(T,\Psi)}_{PE}$.
\end{proof}

Proof of Lemma~\ref{lem:PI_implies_PE}
\begin{proof}\label{App:lem:PI_implies_PE}
Let $\pi_o = \mathcal{U}(\pi, o)$, $i = \mathds{X}_\theta(o)$. 
\begin{align*}
&S^{(T,\Psi)}_{PI}(\pi, \pi_o) \\
\Rightarrow &\Psi(\{\theta: S^*(\pi^\theta, \pi^\theta_o)\})=1 \\
\Rightarrow &\Psi(\{\theta: \forall j \ne i: \pi^\theta_o(i)*s_{R^\theta}(i)>\pi^\theta_o(j)*s_{R^\theta}(j)\})=1
\end{align*}
Let $j_x = \mathds{X}_\theta(x)$. Suppose $S^{(T,\Psi)}_{PE}$ is not satisfied, then $\forall x \ne o: \EX_{\theta \sim \Psi}[\pi^\theta_o(i)*s_{R^\theta}(i)] \le \EX_{\theta \sim \Psi}[\pi^\theta_o(j_x)*s_{R^\theta}(j_x)]$. Then either $\Psi(\{\theta: \forall x \ne o, j_x = i\})=1$, which contradicts the assumption that the partition space $(T, \Psi)$ is point-isolating, or $\Psi(\{\theta: \forall x \ne o: j_x \ne i, \pi^\theta_o(i)*s_{R^\theta}(i) < \pi^\theta_o(j_x)*s_{R^\theta}(j_x)\})>0$.
\end{proof}

\section{Analysis of Update Processes}

\subsection{Update Convergence}

Proof of Lemma~\ref{lem:prior_agnostic}
\begin{proof}\label{App:lem:prior_agnostic}
Assume $\frac{1}{n}\sum_{i=1}^n K_{o_i} \xrightarrow[]{d} X$, then $\forall x$, $\lim_{n \rightarrow \infty} |\frac{1}{n}\sum_{i=1}^n F_{K_{o_i}}(x) - F_X(x)| = 0$. Then $\forall x$:
\begin{align*}
&|\frac{k}{k+n}F_{\pi}(x) + \frac{1}{k+n}\sum_{i=1}^n F_{K_{o_i}}(x) - F_X(x)| \\
\le &|\frac{k}{k+n}F_{\pi}(x)| + |\frac{1}{k+n}\sum_{i=1}^n F_{K_{o_i}}(x) - F_X(x)| \\
\le &|\frac{k}{k+n}F_{\pi}(x)| + |\frac{1}{n}\sum_{i=1}^n F_{K_{o_i}}(x) - F_X(x)|
\end{align*}
We have that $F_\pi(x)$ is bounded in $[0,1]$, so $\lim_{n \rightarrow \infty} |\frac{k}{k+n}F_{\pi}(x)| + |\frac{1}{n}\sum_{i=1}^n F_{K_{o_i}}(x) - F_X(x)| = \lim_{n \rightarrow \infty} |\frac{k}{k+n}F_{\pi}(x)| \le \lim_{n \rightarrow \infty} |\frac{k}{k+n}| = 0$. Therefore, $\frac{k}{k+n}\pi + \frac{1}{k+n}\sum_{i=1}^n K_{o_i} \xrightarrow[]{d} X$.

For the other direction, first we note that $\frac{1}{n}\sum_{i=1}^n F_{K_{o_i}}(x) = \frac{k}{n(k+n)} F_{K_{o_i}}(x) + \frac{1}{k+n} F_{K_{o_i}}(x) \le \frac{k}{n(k+n)} + \frac{1}{k+n} F_{K_{o_i}}(x)$.

Assume $\forall x$, $\lim_{n \rightarrow \infty} |\frac{k}{k+n}F_{\pi}(x) + \frac{1}{k+n}\sum_{i=1}^n F_{K_{o_i}}(x) - F_X(x)| = 0$. Note that $\lim_{n \rightarrow \infty} |\frac{k}{k+n}F_{\pi}(x)| \le \lim_{n \rightarrow \infty} |\frac{k}{k+n}| = 0$, so $\lim_{n \rightarrow \infty} |\frac{1}{k+n}\sum_{i=1}^n F_{K_{o_i}}(x) - F_X(x)| = 0$. Then $\forall x$:
\begin{align*}
&|\frac{1}{n}\sum_{i=1}^n F_{K_{o_i}}(x) - F_X(x)| \\
= &|\frac{k}{n(k+n)} F_{K_{o_i}}(x) + \frac{1}{k+n} F_{K_{o_i}}(x) - F_X(x)| \\
\le &|\frac{k}{n(k+n)} F_{K_{o_i}}(x)| + |\frac{1}{k+n} F_{K_{o_i}}(x) - F_X(x)|
\end{align*}
The limit of this expression is 0, so $\frac{1}{n}\sum_{i=1}^n K_{o_i} \xrightarrow[]{d} X$.
\end{proof}

Proof of Proposition~\ref{prop:EmpUpdate_converge}
\begin{proof}\label{App:prop:EmpUpdate_converge}
From Lemma~\ref{lem:prior_agnostic}, the Empirical Update process converges in distribution to $\mathcal{E}_{\{o_i\}} = \frac{1}{n} \sum_{i=1}^n K_o$, which is the \emph{Empirical Measure}. The convergence in distribution of the Empirical Measure comes directly from the Glivenko-Cantelli theorem \cite{glivenko1933sulla, cantelli1933sulla}.
\end{proof}

Proof of Theorem~\ref{thm:ConcentrationConverge}
\begin{proof}\label{App:thm:ConcentrationConverge}
Let $\mathcal{E}_{O_n}$ be the Empirical Measure. Assuming that $\forall \epsilon>0$, $\lim_{n \rightarrow \infty} \frac{1}{n}\sum_{i=1}^{n} C_i(\epsilon) = 0$, we calculate upper and lower bounds on the partial sums of the CDFs of the kernels $H_n(x)$:
\begin{align*}
H_n(x) &\ge \frac{1}{n} \sum_{i=1}^n \mathds{1}_{\{o_i < x-\epsilon\}}(1 - C_i(\epsilon))\\
&\ge \frac{1}{n} \sum_{i=1}^n \mathds{1}_{\{o_i < x-\epsilon\}} - \frac{1}{n} \sum_{i=1}^{n} C_i(\epsilon)\\
&= F_{\mathcal{E}_{O_n}}(x-\epsilon) - \frac{1}{n} \sum_{i=1}^{n} C_i(\epsilon)\\
\Rightarrow &\liminf_n H_n(x) \ge \liminf_n F_{\mathcal{E}_{O_n}}(x-\epsilon)\\
\Rightarrow &\liminf_n H_n(x) \ge F(x) \text{ at continuity points}\\
\end{align*}
Symmetrically:
\begin{align*}
H_n(x) &\le \frac{1}{n} \sum_{i=1}^n \mathds{1}_{\{o_i > x+\epsilon\}}C_i(\epsilon)\\
&\le \frac{1}{n} \sum_{i=1}^{n} C_i(\epsilon)\\
\Rightarrow 1-H_n(x) &\ge 1 - \frac{1}{n} \sum_{i=1}^{n} C_i(\epsilon)\\
&\ge \frac{1}{n} \sum_{i=1}^n \mathds{1}_{\{o_i > x+\epsilon\}} - \frac{1}{n} \sum_{i=1}^{n} C_i(\epsilon)\\
&= 1 - F_{\mathcal{E}_{O_n}}(x+\epsilon) - \frac{1}{n} \sum_{i=1}^{\infty} C_i(\epsilon)\\
\Rightarrow &\liminf_n (1- H_n(x)) \ge \liminf_n (1- F_{\mathcal{E}_{O_n}}(x+\epsilon))\\
\Rightarrow &\limsup_n H_n(x) \le F(x) \text{ at continuity points}\\
\end{align*}
Therefore $H_n(x) \xrightarrow[]{d} F(x)$.
\end{proof}

Proof of Corollary~\ref{cor:BoundedConvergence}
\begin{proof}\label{App:cor:BoundedConvergence}
Define $X_i$ as the random variables distributed according to $K_{o_i}$ and $Y_i = |X_i - o_i|$. Given any $\epsilon>0$, from the convergence of $\Delta_i$, we have that $\exists N$ such that $\forall n>N$, $\delta_{n,j}<\epsilon$. So $\forall i > N$, $C_i(\epsilon) = 0$. Therefore $\lim_{n \rightarrow \infty} \sum_{i=1}^n C_i(\epsilon) = \sum_{i=1}^N C_i(\epsilon) \Rightarrow \lim_{n \rightarrow \infty} \frac{1}{n}\sum_{i=1}^n C_i(\epsilon) = 0$. From Theorem~\ref{thm:ConcentrationConverge}, the sequence of averages of the CDFs of the kernels $H_n(x) \xrightarrow[]{d} F(x)$. It then follows directly from Lemma~\ref{lem:prior_agnostic} that the update process is convergent.
\end{proof}

\subsection{Satisfying the Update Conditions}

Proof of Theorem~\ref{thm:EmpUpdate_PI}
\begin{proof}\label{App:thm:EmpUpdate_PI}
The Empirical Update process yields $\pi_o = (1-\alpha)*R + \alpha*K_o$. Let $i = \mathds{X}_\theta(o)$. Then $\forall \theta:$ $\pi^\theta_o = (1 - \alpha)*R^\theta + \alpha*K^\theta_o$ where $K^\theta_o(j) = \mathds{1}_{j=i}$. Then $\frac{\pi^\theta_o(j)}{R^\theta(j)} = (1 - \alpha) + \alpha \frac{K^\theta_o(j)}{R^\theta(j)} = \begin{cases}
(1 - \alpha) + \frac{\alpha}{R^\theta(j)} & j=i \\
(1 - \alpha) & j \ne i
\end{cases}$. The assumption that $(T, \Psi)$ is bin-supported over $R$ ensures that $R^\theta(j)>0$. Therefore, $\forall \theta:$ $\forall j \ne i:$ $\frac{\pi^\theta_o(i)}{R^\theta(i)} > \frac{\pi^\theta_o(j)}{R^\theta(j)}$.
\end{proof}

\subsection{Bin Edge Conditions}

\subsubsection{Failing PI}

Proof of Lemma~\ref{lem:BoundedCenterMass}
\begin{proof}\label{App:lem:BoundedCenterMass}
We first note that if two or more bins $\beta_1$ and $\beta_2$ must have equal ratios of posterior to prior, it must have equal ratios of kernel to prior: 
\begin{align*}
\frac{\pi_o(\beta_1)}{\pi(\beta_1)} &= \frac{\pi_o(\beta_2)}{\pi(\beta_2)} \\
\implies (1 - \alpha) + \alpha \frac{K_o(\beta_1)}{\pi(\beta_1)} &= (1 - \alpha) + \alpha \frac{K_o(\beta_2)}{\pi(\beta_2)} \\
\implies \frac{K_o(\beta_1)}{\pi(\beta_1)} &= \frac{K_o(\beta_2)}{\pi(\beta_2)}
\end{align*}
If we place the corner of four bins on the observation point $z$, let the prior probabilities of the four bins be $A_{r,u}, A_{r,b}, A_{l,u}, A_{l,b}$ corresponding to the upper-right, bottom-right, upper-left, and bottom-left corners. Consider also a bin boundary on $o$ such that the left bin is $\beta_l = [(o_1 - l_1, o_2 - \frac{l_2}{2}), (o_1, o_2 + \frac{l_2}{2}))$ and the right bin is $\beta_r = [(o_1, o_2 - \frac{l_2}{2}), (o_1 + l_1, o_2 + \frac{l_2}{2}))$. Consider a prior such that $\pi(\beta_l) = A_{l,u} + A_{l,b}$ and $\pi(\beta_r) = r(A_{r,u} + A_{r,b})$ where we can construct the prior so $r$ takes on any value in $[0, 1]$.

Now consider the probabilities of the kernel in the four corners of $A$: $A^*_{r,u}, A^*_{r,b}, A^*_{l,u}, A^*_{l,b}$. PI requires that $(A^*_{r,u}, A^*_{r,b}, A^*_{l,u}, A^*_{l,b}) = \lambda_1 (A_{r,u}, A_{r,b}, A_{l,u}, A_{l,b})$. We apply the same PI constraint to the centered left and right bins: $K_o(\beta_l \cap A) = \lambda_2 (A_{l,u} + A_{l,b})$ and $K_o(\beta_r \cap A) = \lambda_2 r(A_{r,u} + A_{r,b})$ with $\lambda_2 < \lambda_1$. Suppose that at most $1-x$ fraction of the kernel probability is outside $A$. Then we have the following inequalities:
\begin{align*}
(1-x) \lambda_1 (A_{l,u} + A_{l,b}) &\ge (\lambda_1 - \lambda_2) (A_{l,u} + A_{l,b}) \ge 0 \\
(1-x) \lambda_1 (A_{r,u} + A_{r,b}) &\ge (\lambda_1 - \lambda_2 r) (A_{r,u} + A_{r,b}) \ge 0 \\
\implies (1-x) &\ge 1 - \frac{\lambda_2}{\lambda_1}r > 0 \\
\implies x &\le \frac{\lambda_2}{\lambda_1}r < 1
\end{align*}
\end{proof}

Proof of Theorem~\ref{thm:PI_impossible}
\begin{proof}\label{App:thm:PI_impossible}
From Lemma~\ref{lem:BoundedCenterMass}, there is a prior $\pi$ such that, in order to satisfy PI, there exists a $\Delta$ such that if $A = [z - \Delta, z + \Delta]$, the kernel value $K_o(A)$ is uniformly bounded above. This uniform bound is itself bounded above by $r<1$, defined in Lemma~\ref{lem:BoundedCenterMass} as the ratio of the prior probabilities in the two bins to the right of $o$ to the prior probabilities in the two bins to the left of $o$. In order for the additive update with $K_o$ to satisfy PI, this ratio $r$ must be invariant with respect to the additive update sequence. Therefore, if the Agent wishes to update over a sequence of observations $O_n$ with a sequence of kernels $\{K_{o_i}\}$ such that $\pi_i = \frac{k+i-1}{k+i} \pi_{i-1} + \frac{1}{k+i} K_{o_i}$, then all the kernels have the value $K_{o_i}(A)$ uniformly bounded above by $r$. Then $\lim_{n \rightarrow \infty} \pi_n(A) \le \max_{i=1}^{\infty} K_{o_i}(A) \le r<1$. If the true distribution $\Phi$ is more heavily concentrated inside $A$, such that $\Phi(A) = r_{\Phi} > r$. Then $\lim_{n \rightarrow \infty} |F_{\pi_n}(x) - F_{\Phi}(x)|$ is uniformly bounded below by $r_{\Phi} - r > 0$. Therefore, this update sequence cannot satisfy the condition in Lemma~\ref{thm:ConcentrationConverge}, and therefore cannot be convergent.
\end{proof}

\subsubsection{Satisfying PE}

Proof of Theorem~\ref{thm:PEPyramids}
\begin{proof}\label{App:thm:PEPyramids}
We are given the following bin boundary conditions for satisfying PE:
\begin{align*}
&\int_{-\frac{L_{-i}}{2}}^{\frac{L_i}{2}} Q_o(\delta_1, \dots , -\frac{l_i}{2}, \dots , \delta_d) \partial \Delta_i \\
= &\int_{-\frac{L_{-i}}{2}}^{\frac{L_i}{2}} Q_o(\delta_1, \dots , \frac{l_i}{2}, \dots , \delta_d) \partial \Delta_i
\end{align*}

We will construct the kernel $K^x_o$ as having a PDF that is a pyramid with the peak at $o$ and the base at $[x-\Delta, x+\Delta]$ with $\Delta < L$ the dimensions of the bins. We prove that there exists an $x \in [o - \Delta, o + \Delta]$ such that the kernel satisfies PI for the PTNE mechanism. We will demonstrate the construction on $\mathbb{R}^2$, but the argument is applicable to all dimensions.

Define $Q_x(\omega) = \dfrac{\widetilde{f}_{K^x_o}(\omega)}{\widetilde{f}_{R}(\omega)}$. We define $S(x)$ as the integrals of $Q_x(\omega)$ over the four edges of the rectangle $[o - \frac{L}{2}, o + \frac{L}{2}]$, with $x$ being the location of the center of the base of the pyramid:
\begin{align*}
S_l(x) &= \int_{-\frac{l_2}{2}}^{\frac{l_2}{2}} Q_x(o + \langle -\frac{l_1}{2}, y_2 \rangle) \partial y_2 \\
S_r(x) &= \int_{-\frac{l_2}{2}}^{\frac{l_2}{2}} Q_x(o + \langle \frac{l_1}{2}, y_2 \rangle) \partial y_2 \\
S_b(x) &= \int_{-\frac{l_1}{2}}^{\frac{l_1}{2}} Q_x(o + \langle y_1, -\frac{l_2}{2} \rangle) \partial y_1 \\
S_u(x) &= \int_{-\frac{l_1}{2}}^{\frac{l_1}{2}} Q_x(o + \langle y_1, \frac{l_2}{2} \rangle) \partial y_1
\end{align*}
To satisfy PE, according to the bin boundary conditions, we must find an $x$ such that $S_l(x) = S_r(x)$ and $S_b(x) = S_u(x)$. Define $F_h(x) = S_r(x) - S_l(x)$ and $F_v(x) = S_u(x) - S_b(x)$ as the horizontal and vertical residuals. We observe that when $x_1 = o_1 - \delta_1$, $F_h < 0$, and when $x_1 = o_1 + \delta_1$, $F_h > 0$. Similarly, when $x_2 = o_2 - \delta_2$, $F_v < 0$, and when $x_2 = o_2 + \delta_2$, $F_v > 0$. The two functions $F_h(x)$ and $F_v(x)$ satisfy the assumptions laid out in the Poincaré-Miranda Theorem on the box $[o-\Delta, o+\Delta]$ \cite{miranda1940osservazione}. Therefore, there exists an $x$ in the box such that $F_h(x)=0$ and $F_v(x)=0$, thus satisfying the PE condition.

In higher dimensions, the four $S$ functions simply correspond to integrals of $Q$ over the faces of the rectangular hyper-prism. We can define functions $F_i$ corresponding to the residuals in $S$ on opposing faces in the coordinate direction $i$, and the Poincaré-Miranda Theorem applies as before.

We can take this a step further, because we know that $F_i(x)$ is monotonic in $x_i$, so the function $G(x) = \sum_{i=1}^d F^2_i(x)$ is guaranteed to converge to a solution $x$ via gradient descent. So such a continuous update kernel then not only exists, but is computable.
\end{proof}

\section{Simulations}\label{App:Simulations}

\subsection{Report Perturbation}
To generate the distributions, we sample 5 values uniformly in $[0, 1)$ for 1D and $[0, 1) \times [0, 1)$ for 2D. For both the True and Public Distributions, each value is weighted with an independent random variable in $[0, 1)$, and the weight vector is normalized. The confidence value for the update is 1. The bin size is $0.2$ for 1D and $\sqrt{0.2} \times \sqrt{0.2}$ for 2D. The Partition Selection is just a translation by a random variable sampled uniformly from a bin volume.

\paragraph{Empirical Distribution, Empirical Update}
Values and weights are treated as weighted delta functions. Expectations are taken over 200 Peer reports, 500 Partition Selection samples for 1D and 400 samples for 2D. Perturbations go from $-1$ to $1$ in intervals of $\frac{1}{30}$ for 1D, and from $(-1, -1)$ to $(1, 1)$ in intervals of $\frac{1}{10} \times \frac{1}{10}$ for 2D.

\begin{figure}[ht]
    \centering
    \includegraphics[width=0.84\linewidth]{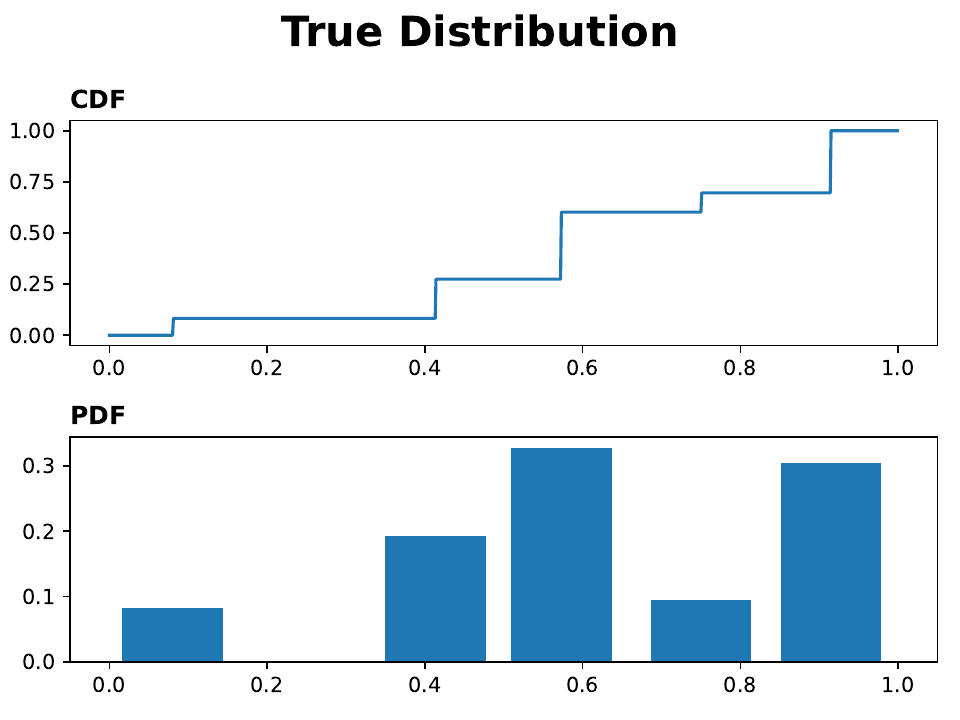}
    \includegraphics[width=0.84\linewidth]{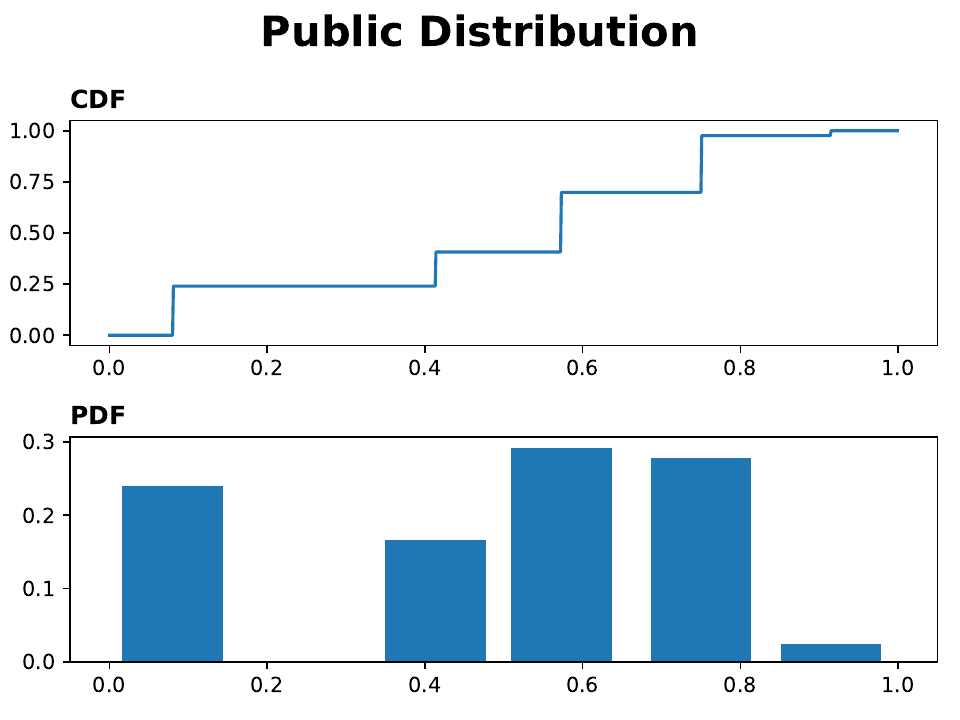}
    \includegraphics[width=0.84\linewidth]{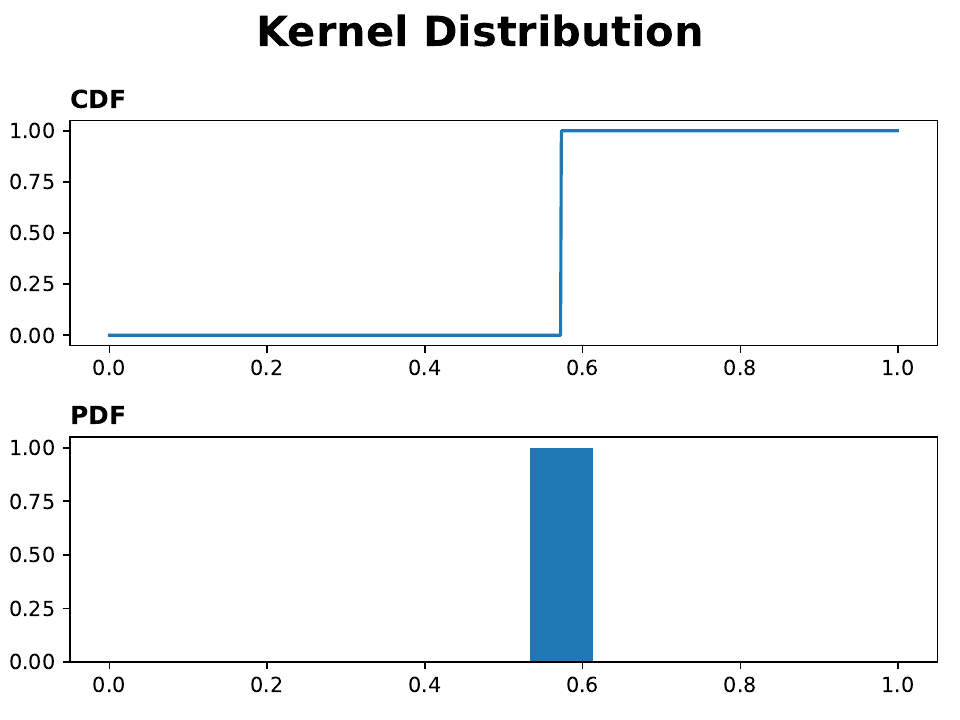}
    \includegraphics[width=0.84\linewidth]{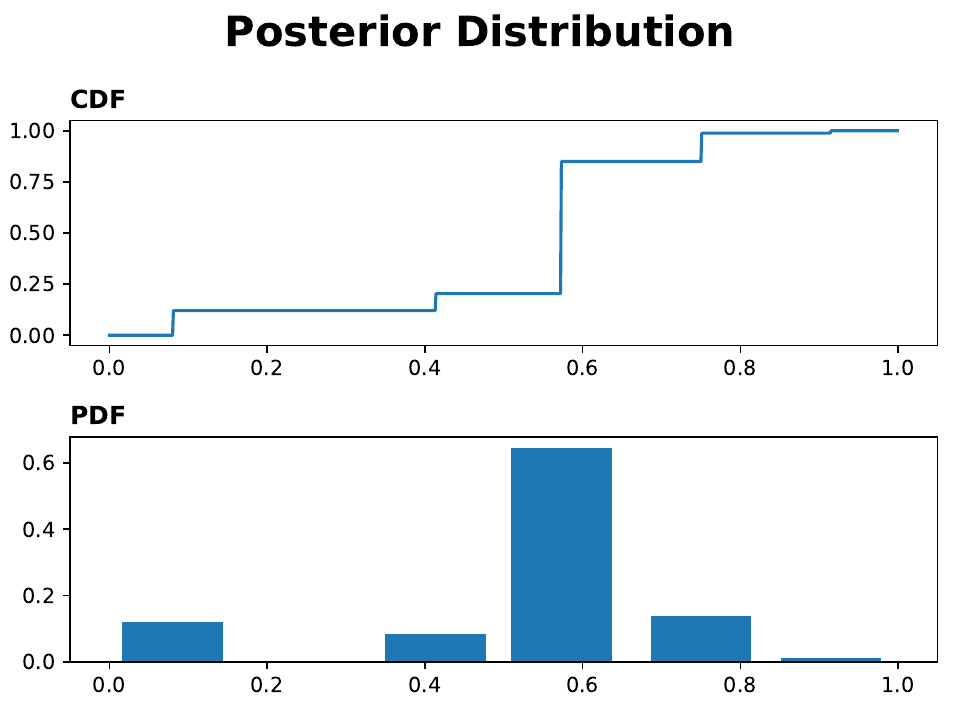}
    \caption{True, Public, Kernel, and Posterior distributions for 1D Empirical distribution, Empirical update perturbation simulations.}
    \label{fig:empirical_1d_perturbations}
\end{figure}

\begin{figure}[ht]
    \centering
    \includegraphics[width=0.99\linewidth]{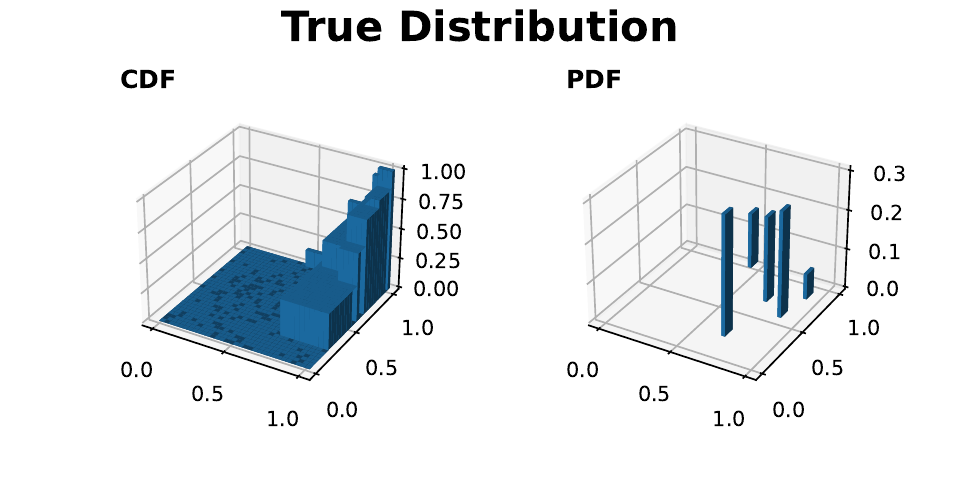}
    \includegraphics[width=0.99\linewidth]{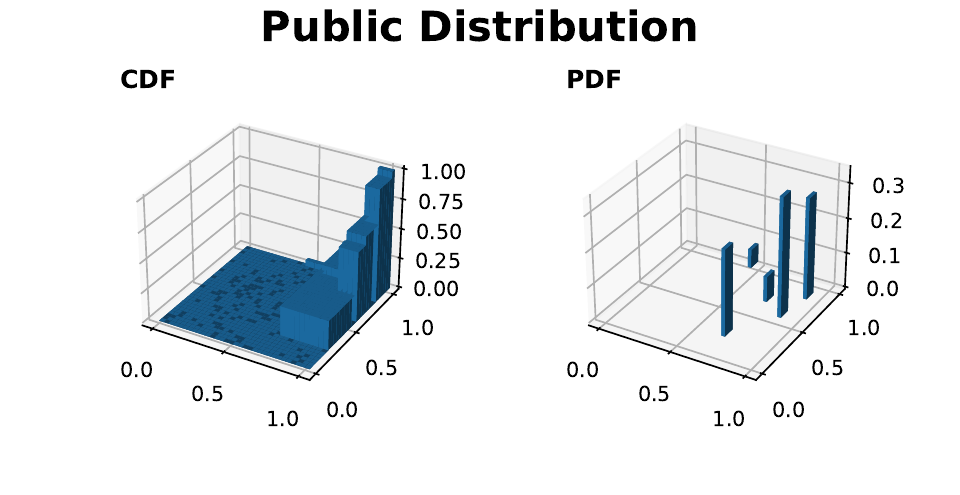}
    \includegraphics[width=0.99\linewidth]{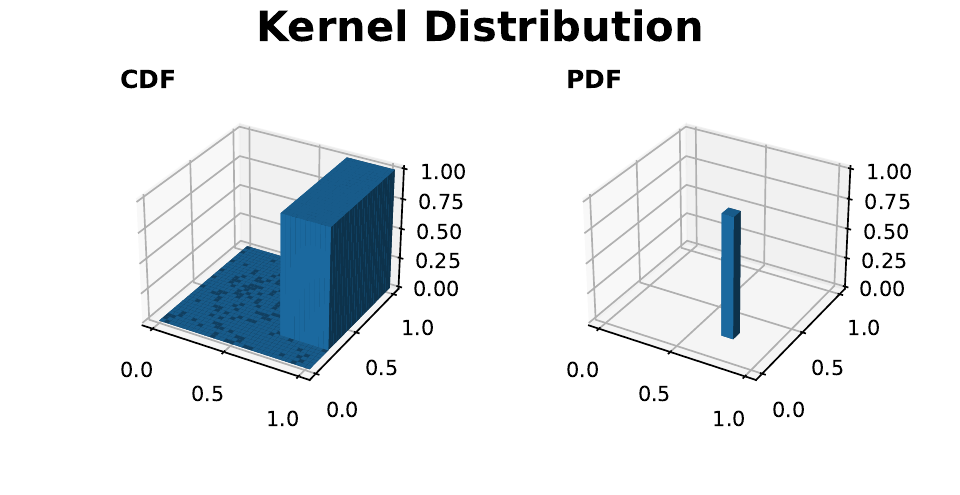}
    \includegraphics[width=0.99\linewidth]{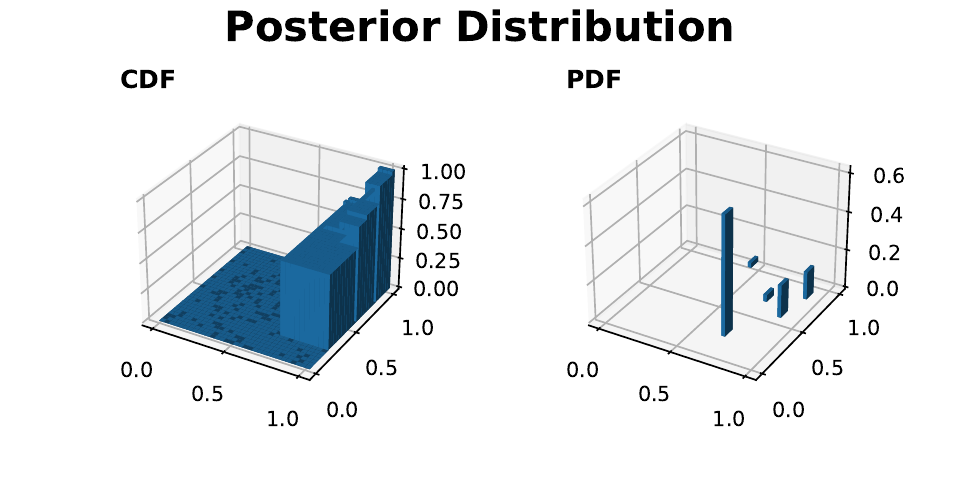}
    \caption{True, Public, Kernel, and Posterior distributions for 2D Empirical distribution, Empirical update perturbation simulations.}
    \label{fig:empirical_2d_perturbations}
\end{figure}

\paragraph{GMM Distribution, Pyramid Update}
Values are treated as means of Gaussian distributions. The variance in 1D is taken as $2*\min_{i \ne j}(|V_i - V_j|)$ where $V_i$ and $V_j$ are from the value list. The covariance in 2D is taken as a diagonal matrix with $2*\min_{i \ne j}(|V_i - V_j|)$ for each coordinate. The size of the Pyramid kernel base is the one tenth the bin size. Expectations are taken over 200 Peer reports, 200 Partition Selection samples for 1D and 64 samples for 2D. Perturbations go from $-1$ to $1$ in intervals of $\frac{1}{30}$ for 1D, and from $(-1, -1)$ to $(1, 1)$ in intervals of $\frac{1}{8} \times \frac{1}{8}$ for 2D.

\begin{figure}[ht]
    \centering
    \includegraphics[width=0.84\linewidth]{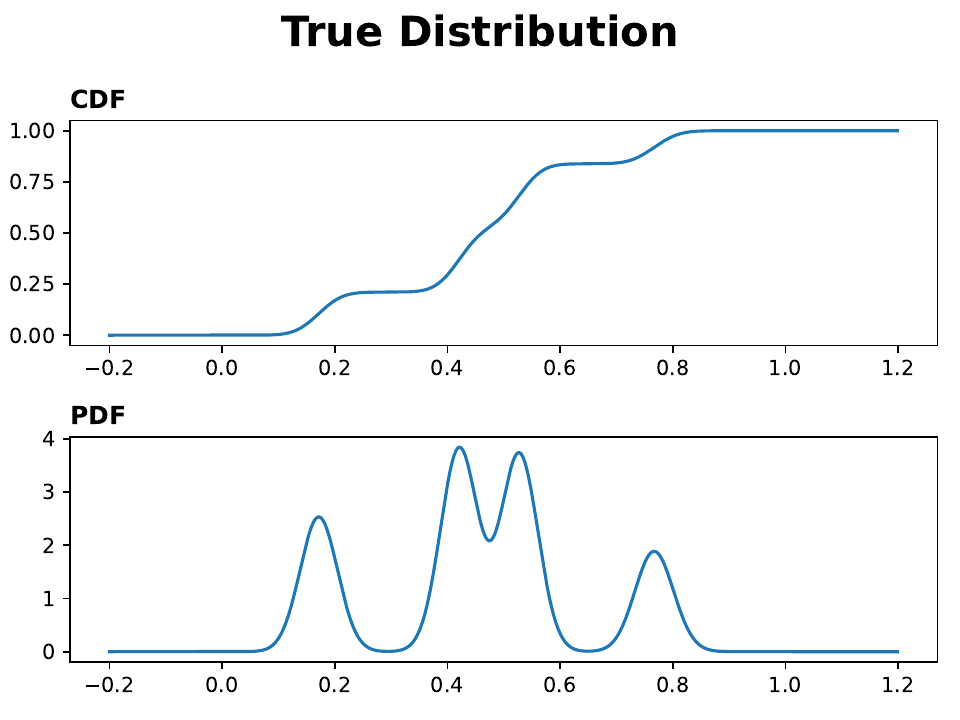}
    \includegraphics[width=0.84\linewidth]{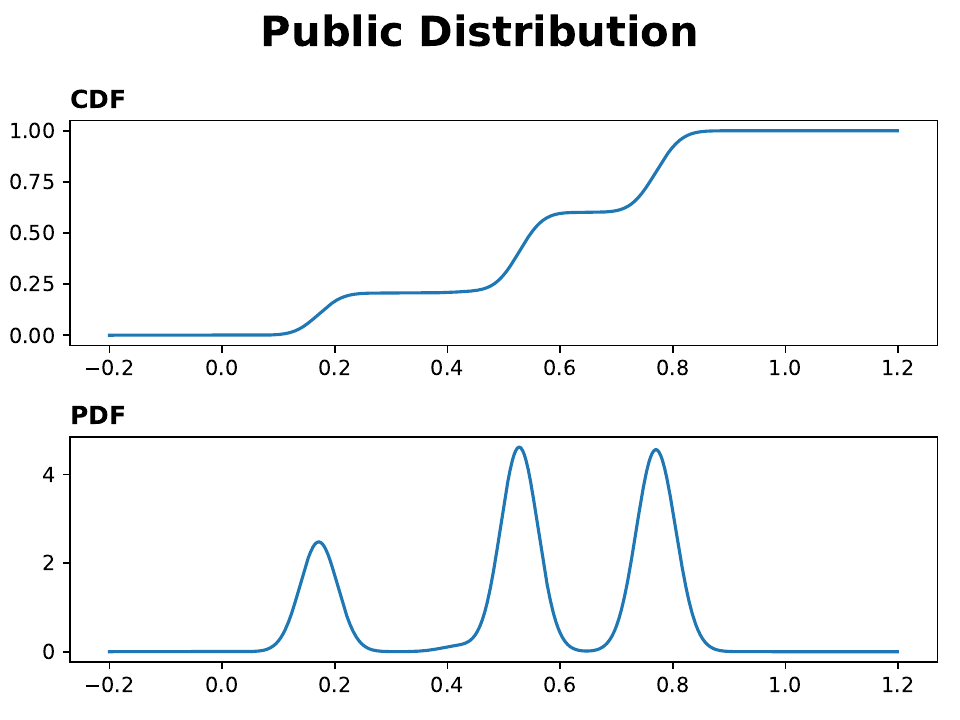}
    \includegraphics[width=0.84\linewidth]{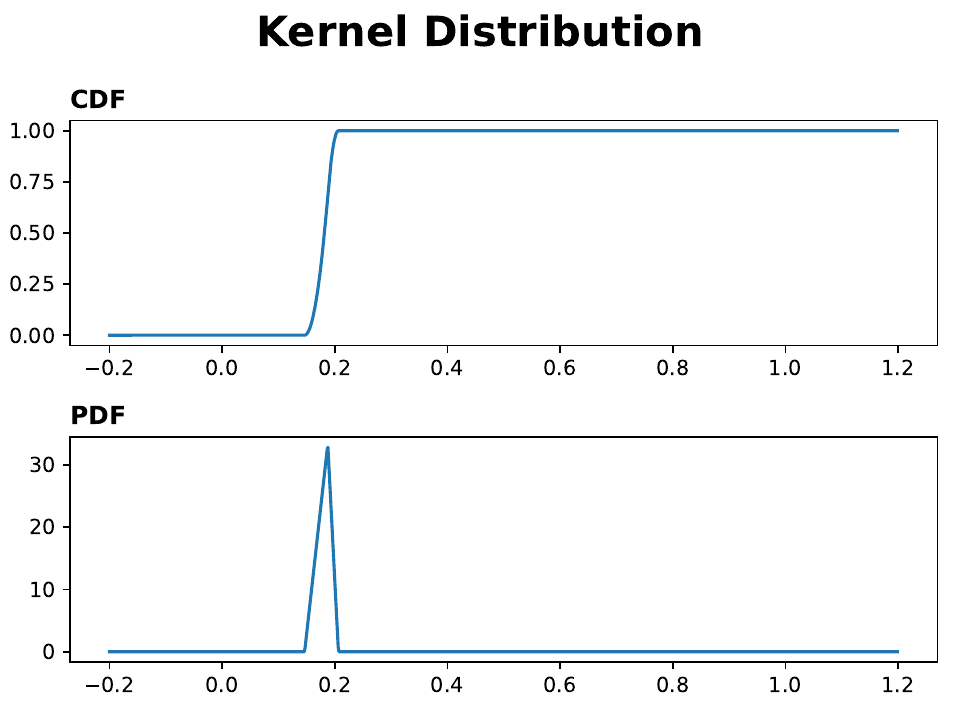}
    \includegraphics[width=0.84\linewidth]{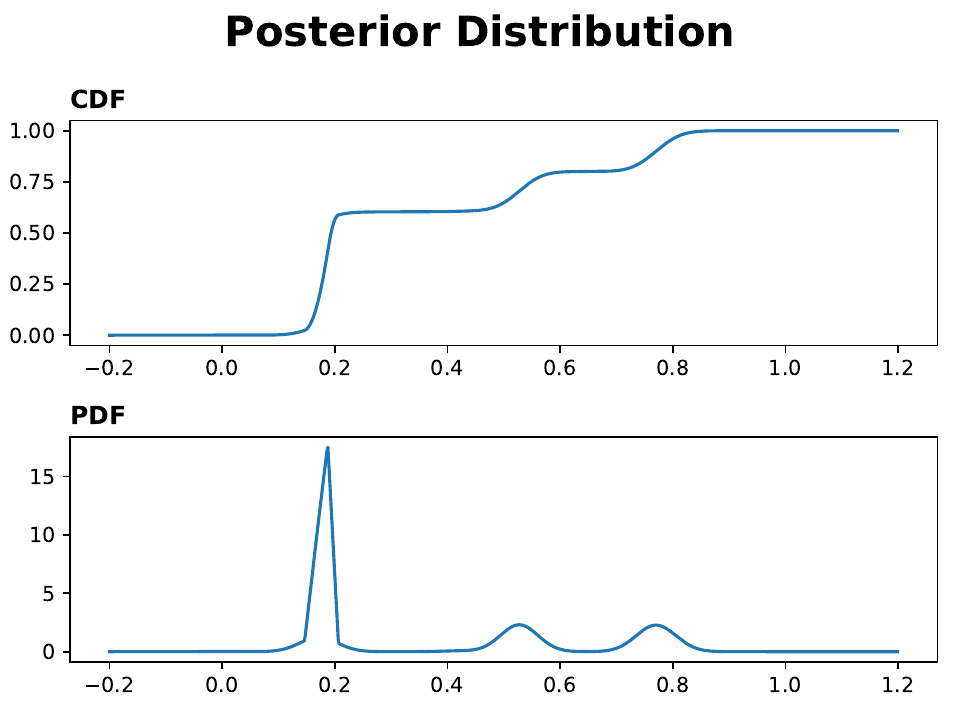}
    \caption{True, Public, Kernel, and Posterior distributions for 1D GMM distribution, Pyramid update perturbation simulations.}
    \label{fig:gauss_1d_perturbations}
\end{figure}

\begin{figure}[ht]
    \centering
    \includegraphics[width=0.99\linewidth]{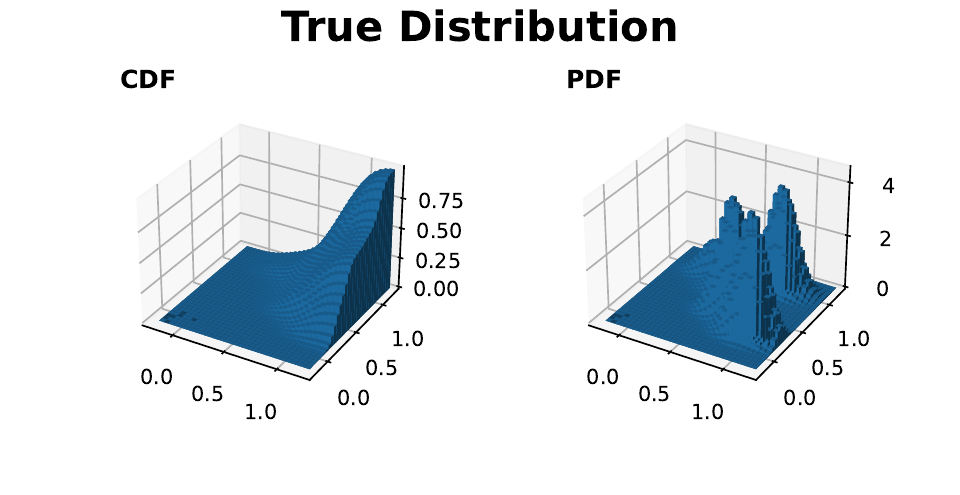}
    \includegraphics[width=0.99\linewidth]{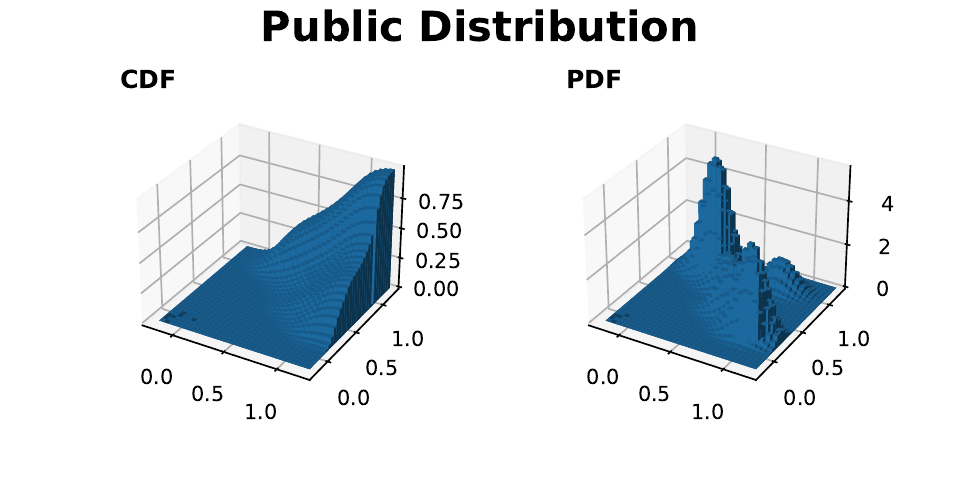}
    \includegraphics[width=0.99\linewidth]{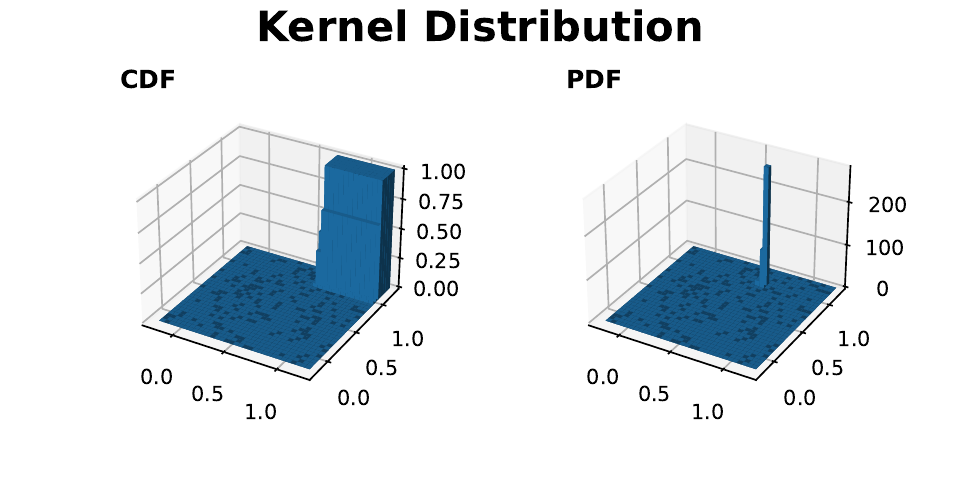}
    \includegraphics[width=0.99\linewidth]{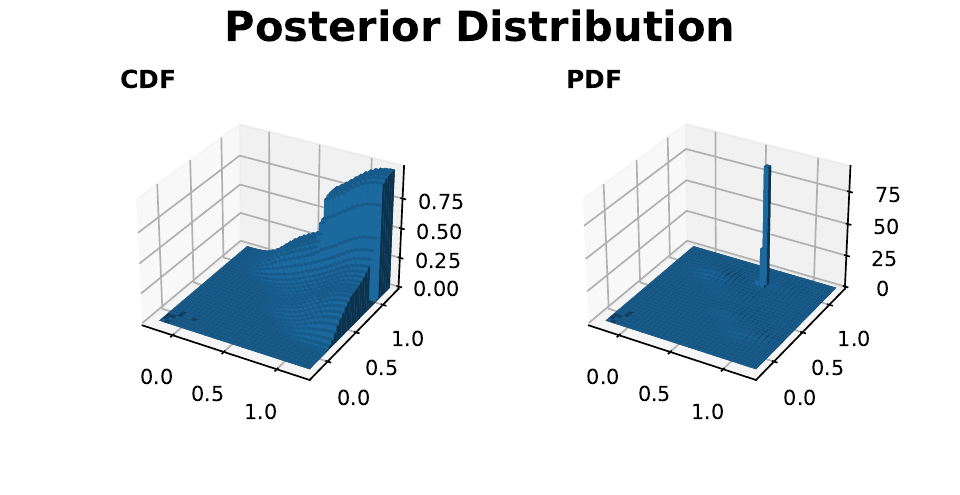}
    \caption{True, Public, Kernel, and Posterior distributions for 2D GMM distribution, Pyramid update perturbation simulations.}
    \label{fig:gauss_2d_perturbations}
\end{figure}

\subsection{Payment Stability}
The distributions are all generated the same way as in the previous section, but with the bin size varying from $\frac{1}{30}$ to $1$ in intervals of $\frac{1}{30}$.

\begin{figure}[ht]
    \centering
    \includegraphics[width=0.82\linewidth]{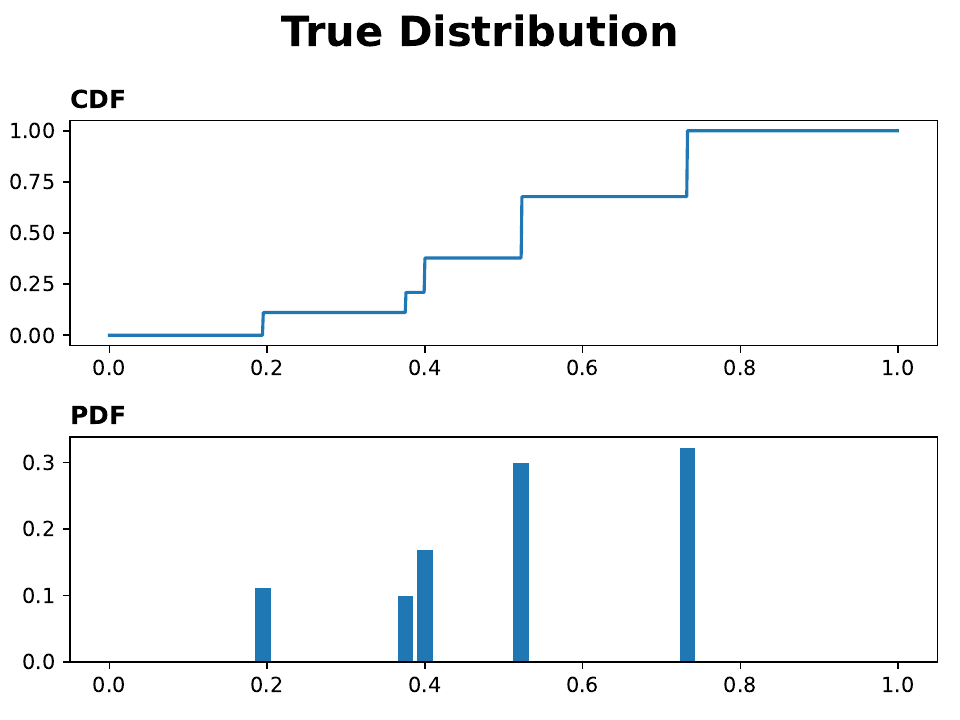}
    \includegraphics[width=0.82\linewidth]{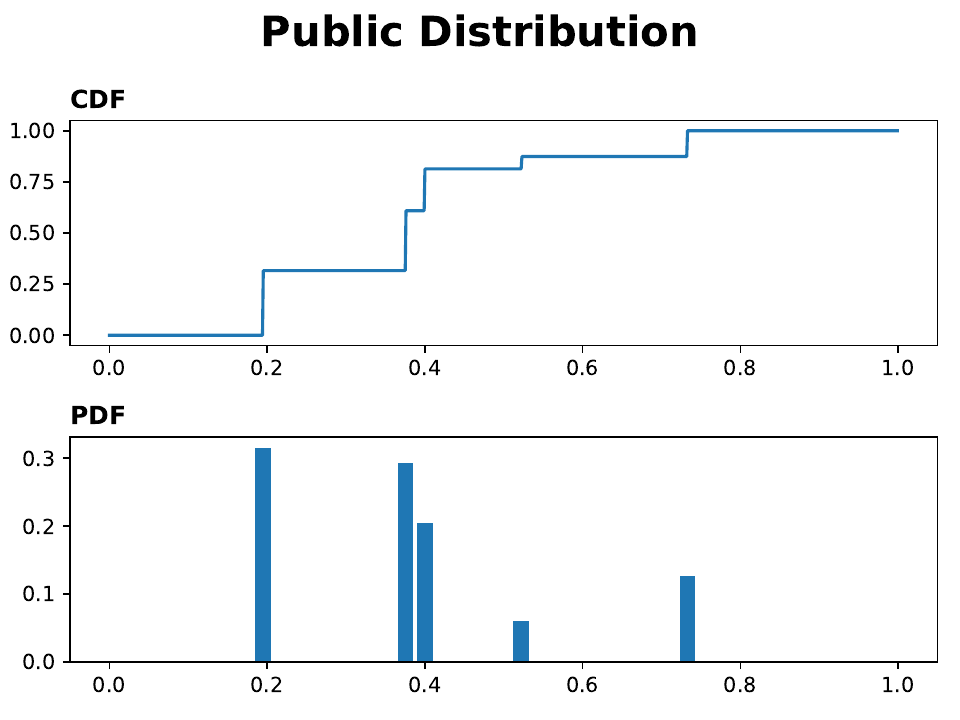}
    \includegraphics[width=0.82\linewidth]{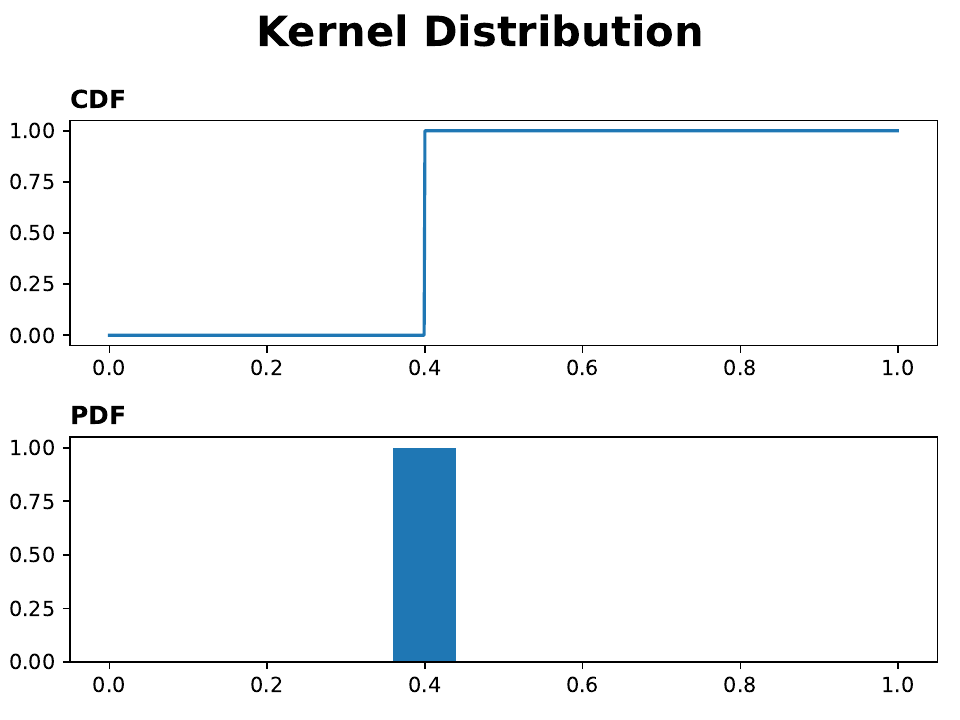}
    \includegraphics[width=0.82\linewidth]{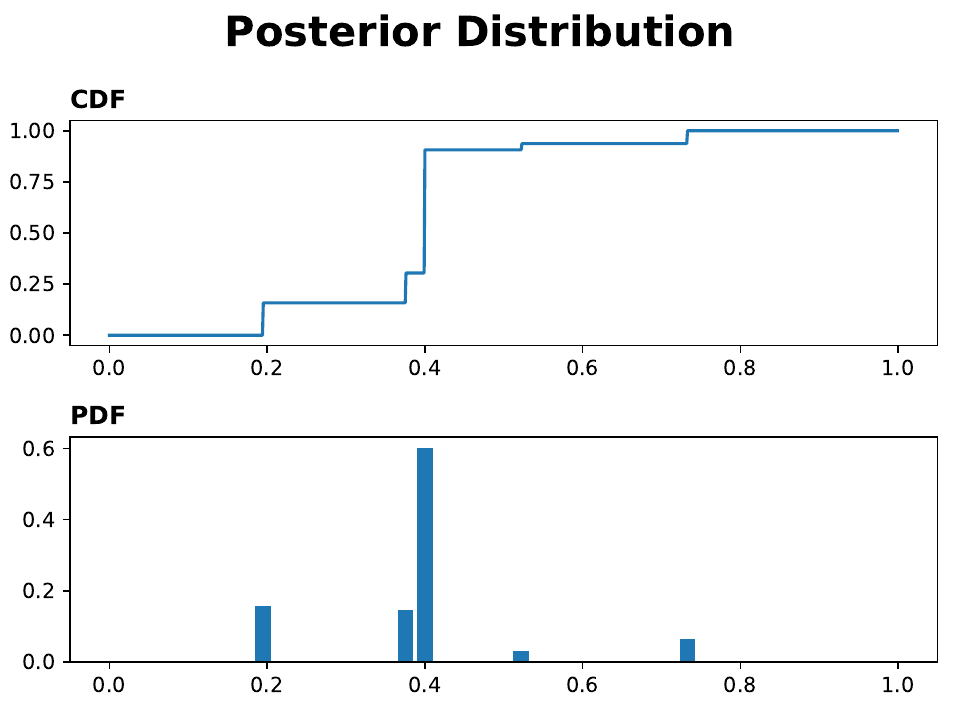}
    \caption{True, Public, Kernel, and Posterior distributions for Empirical distribution, Empirical update bin size simulations. The Kernel and Posterior distributions are taken with the largest bin size.}
    \label{fig:empirical_1d_boxes}
\end{figure}

\begin{figure}[ht]
    \centering
    \includegraphics[width=0.82\linewidth]{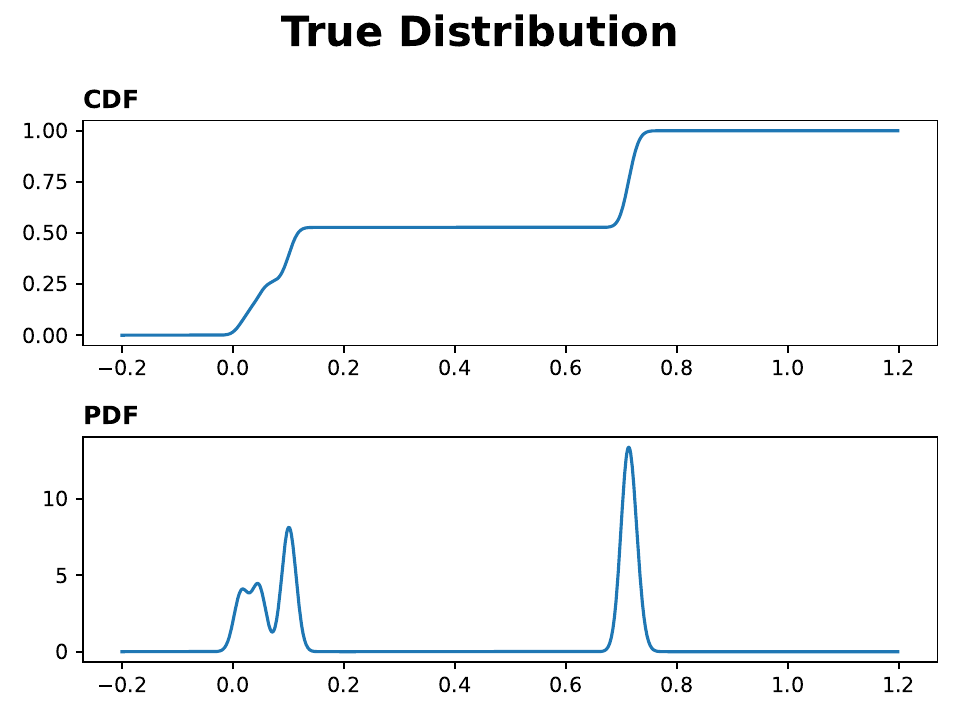}
    \includegraphics[width=0.82\linewidth]{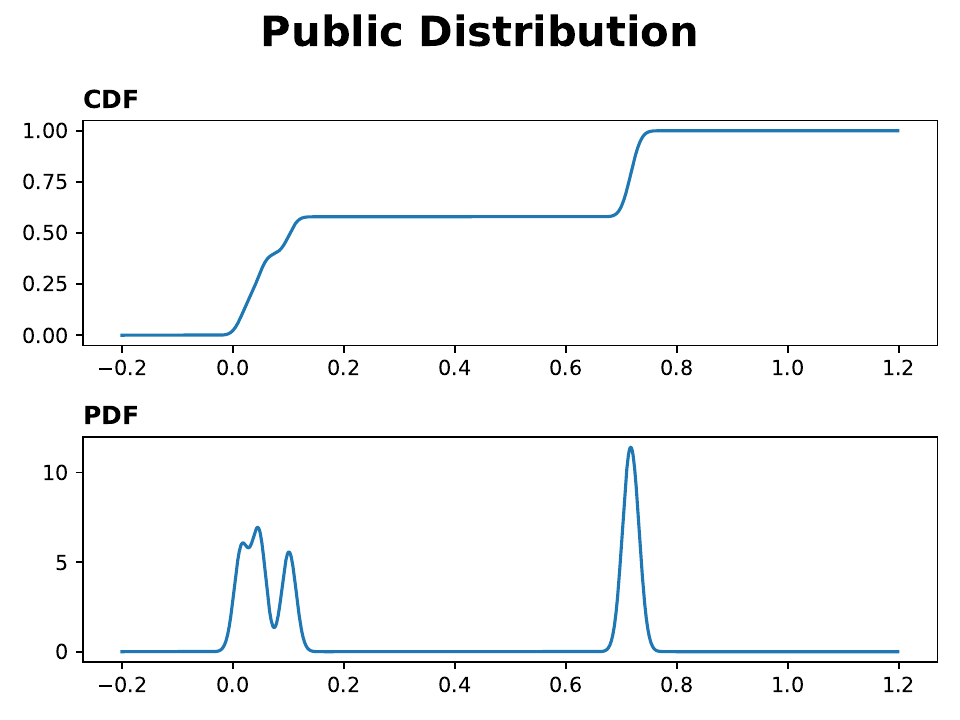}
    \includegraphics[width=0.82\linewidth]{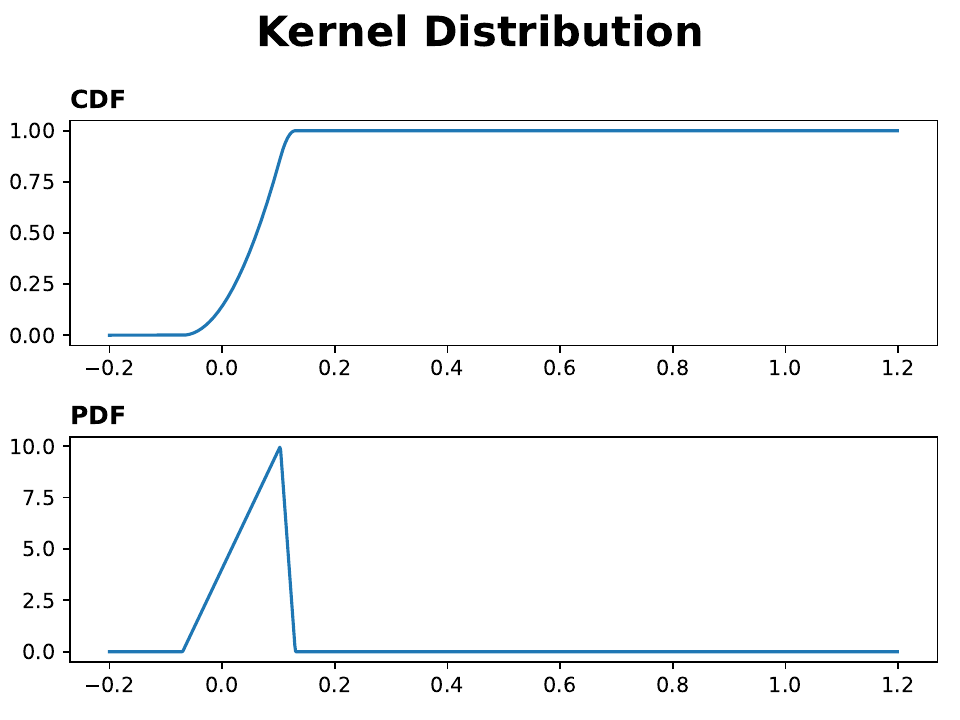}
    \includegraphics[width=0.82\linewidth]{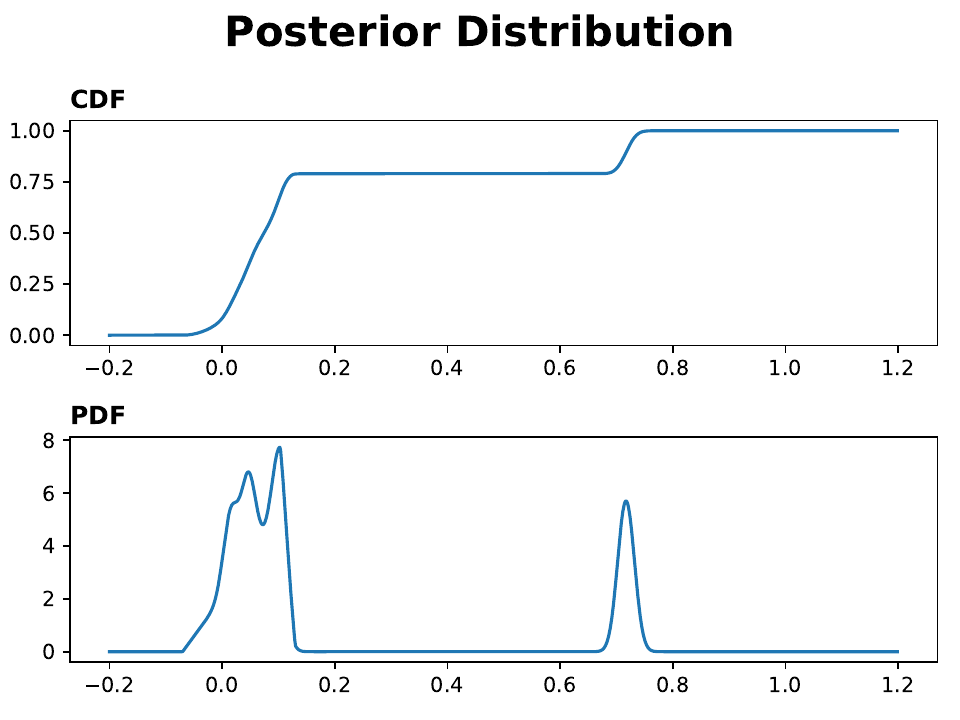}
    \caption{True, Public, Kernel, and Posterior distributions for GMM distribution, Pyramid update bin size simulations. The Kernel and Posterior distributions are taken with the largest bin size.}
    \label{fig:gauss_1d_boxes}
\end{figure}

\bibliography{references}

\begin{thebibliography}{19}
\providecommand{\natexlab}[1]{#1}

\bibitem[{Cantelli(1933)}]{cantelli1933sulla}
Cantelli, F.~P. 1933.
\newblock Sulla determinazione empirica delle leggi di probabilita.
\newblock \emph{Giorn. Ist. Ital. Attuari}, 4(421-424).

\bibitem[{Chen, Shen, and Zheng(2020)}]{chen2020truthful}
Chen, Y.; Shen, Y.; and Zheng, S. 2020.
\newblock Truthful data acquisition via peer prediction.
\newblock \emph{Advances in Neural Information Processing Systems}, 33: 18194--18204.

\bibitem[{Dasgupta and Ghosh(2013)}]{dasgupta2013crowdsourced}
Dasgupta, A.; and Ghosh, A. 2013.
\newblock Crowdsourced judgement elicitation with endogenous proficiency.
\newblock In \emph{Proceedings of the 22nd international conference on World Wide Web}, 319--330.

\bibitem[{d'Aspremont and G{\'e}rard-Varet(1979)}]{d1979incentives}
d'Aspremont, C.; and G{\'e}rard-Varet, L.-A. 1979.
\newblock Incentives and incomplete information.
\newblock \emph{Journal of Public economics}, 11(1): 25--45.

\bibitem[{Glivenko(1933)}]{glivenko1933sulla}
Glivenko, V. 1933.
\newblock Sulla determinazione empirica delle leggi di probabilita.
\newblock \emph{Gion. Ist. Ital. Attauri.}, 4: 92--99.

\bibitem[{Goel and Faltings(2020)}]{goel2020personalized}
Goel, N.; and Faltings, B. 2020.
\newblock Personalized peer truth serum for eliciting multi-attribute personal data.
\newblock In \emph{Uncertainty in Artificial Intelligence}, 18--27. PMLR.

\bibitem[{Kong and Schoenebeck(2019)}]{kong2019information}
Kong, Y.; and Schoenebeck, G. 2019.
\newblock An information theoretic framework for designing information elicitation mechanisms that reward truth-telling.
\newblock \emph{ACM Transactions on Economics and Computation (TEAC)}, 7(1): 1--33.

\bibitem[{McAfee and Reny(1992)}]{mcafee1992correlated}
McAfee, R.~P.; and Reny, P.~J. 1992.
\newblock Correlated information and mecanism design.
\newblock \emph{Econometrica: Journal of the Econometric Society}, 395--421.

\bibitem[{Miller, Resnick, and Zeckhauser(2005)}]{miller2005eliciting}
Miller, N.; Resnick, P.; and Zeckhauser, R. 2005.
\newblock Eliciting informative feedback: The peer-prediction method.
\newblock \emph{Management Science}, 51(9): 1359--1373.

\bibitem[{Miranda(1940)}]{miranda1940osservazione}
Miranda, C. 1940.
\newblock \emph{Un'osservazione su un teorema di Brouwer}.
\newblock Consiglio Nazionale delle Ricerche.

\bibitem[{Prelec(2004)}]{prelec2004bayesian}
Prelec, D. 2004.
\newblock A Bayesian truth serum for subjective data.
\newblock \emph{science}, 306(5695): 462--466.

\bibitem[{Radanovic and Faltings(2014)}]{radanovic2014incentives}
Radanovic, G.; and Faltings, B. 2014.
\newblock Incentives for truthful information elicitation of continuous signals.
\newblock In \emph{Proceedings of the AAAI Conference on Artificial Intelligence}, volume~28.

\bibitem[{Radanovic and Faltings(2015)}]{radanovic2015incentivizing}
Radanovic, G.; and Faltings, B. 2015.
\newblock Incentivizing truthful responses with the logarithmic peer truth serum.
\newblock In \emph{Adjunct Proceedings of the 2015 ACM International Joint Conference on Pervasive and Ubiquitous Computing and Proceedings of the 2015 ACM International Symposium on Wearable Computers}, 1349--1354.

\bibitem[{Radanovic, Faltings, and Jurca(2016)}]{radanovic2016incentives}
Radanovic, G.; Faltings, B.; and Jurca, R. 2016.
\newblock Incentives for effort in crowdsourcing using the peer truth serum.
\newblock \emph{ACM Transactions on Intelligent Systems and Technology (TIST)}, 7(4): 1--28.

\bibitem[{Shnayder et~al.(2016)Shnayder, Agarwal, Frongillo, and Parkes}]{shnayder2016informed}
Shnayder, V.; Agarwal, A.; Frongillo, R.; and Parkes, D.~C. 2016.
\newblock Informed truthfulness in multi-task peer prediction.
\newblock In \emph{Proceedings of the 2016 ACM Conference on Economics and Computation}, 179--196.

\bibitem[{Von~Ahn and Dabbish(2004)}]{von2004labeling}
Von~Ahn, L.; and Dabbish, L. 2004.
\newblock Labeling images with a computer game.
\newblock In \emph{Proceedings of the SIGCHI conference on Human factors in computing systems}, 319--326.

\bibitem[{Waggoner and Chen(2014)}]{waggoner2014output}
Waggoner, B.; and Chen, Y. 2014.
\newblock Output agreement mechanisms and common knowledge.
\newblock In \emph{Proceedings of the AAAI Conference on Human Computation and Crowdsourcing}, volume~2, 220--226.

\bibitem[{Winkler et~al.(1996)Winkler, Munoz, Cervera, Bernardo, Blattenberger, Kadane, Lindley, Murphy, Oliver, and R{\'\i}os-Insua}]{winkler1996scoring}
Winkler, R.~L.; Munoz, J.; Cervera, J.~L.; Bernardo, J.~M.; Blattenberger, G.; Kadane, J.~B.; Lindley, D.~V.; Murphy, A.~H.; Oliver, R.~M.; and R{\'\i}os-Insua, D. 1996.
\newblock Scoring rules and the evaluation of probabilities.
\newblock \emph{Test}, 5(1): 1--60.

\bibitem[{Witkowski and Parkes(2012)}]{witkowski2012robust}
Witkowski, J.; and Parkes, D. 2012.
\newblock A robust bayesian truth serum for small populations.
\newblock In \emph{Proceedings of the AAAI Conference on Artificial Intelligence}, volume~26, 1492--1498.

\end{thebibliography}

\end{document}